\documentclass[twocolumn,tighten]{aastex63}
\usepackage{amsmath}
\usepackage{enumerate}
\usepackage[shortlabels]{enumitem}
\setlist[enumerate]{nosep,nolistsep}
\usepackage{multirow}
\setlist[itemize]{nosep,leftmargin=*}

\graphicspath{{Figures/}}

\usepackage{booktabs}

\newcommand{\blos}[1]{{$\Delta{B}_{{\rm LOS}}$#1}}
\newcommand{\bt}[1]{{$B_{{\rm LOS}}(t)$#1}}
\newcommand{\BLOS}[1]{{\Delta{\rm B}_{{\rm LOS}}#1}}

\newcommand{\ergs}[1]{erg\,s$^{-1}$\,cm$^{-2}$\,sr$^{-1}$\,\AA{}$^{-1}$}

\usepackage[makeroom]{cancel}

\definecolor{darkgreen}{rgb}{0.0, 0.4, 0.13}

\shorttitle{White-light emission and photospheric magnetic field changes in flares}
\shortauthors{Castellanos Dur\'{a}n \& Kleint}
\received{}
\revised{}
\accepted{}
\submitjournal{ApJ}

\begin{document}

\title{The Statistical Relationship between White-light Emission and Photospheric Magnetic Field Changes in Flares}

\correspondingauthor{J. S. Castellanos Dur\'an}
\email{castellanos@mps.mpg.de}

\author[0000-0003-4319-2009]{J. S. Castellanos Dur\'an}
\affiliation{Max Planck Institute for Solar System Research, Justus-von-Liebig-Weg 3, D-37077 G\"ottingen, Germany}
\affiliation{Observatorio Astron\'omico Nacional, Universidad Nacional de Colombia, Carrera 45 No. 26 85, 11001 Bogot\'a, Colombia.}

\author[0000-0002-7791-3241]{Lucia Kleint}
\affiliation{Leibniz-Institut f\"ur Sonnenphysik (KIS), Sch\"oneckstrasse 6, D-79104 Freiburg, Germany.}

\begin{abstract}

Continuum emission, also called white-light emission (WLE), and permanent changes of the magnetic field ($\Delta{B}_{{\rm{LOS}}}$) are often observed during solar flares. But their relation and their precise mechanisms are still unknown. We study statistically the relationship between $\Delta{B}_{{\rm{LOS}}}$ and WLE during 75 solar flares of different strengths and locations on the solar disk. We analyze SDO/HMI data and determine for each pixel in each flare if it exhibited WLE and/or $\Delta{B}_{{\rm{LOS}}}$. We then investigate the occurrence, strength, and spatial size of the WLE, its dependence on flare energy, and its correlation to the occurrence of $\Delta{B}_{{\rm{LOS}}}$. We detected WLE in 44/75 flares and $\Delta{B}_{{\rm{LOS}}}$ in 59/75 flares. We find that WLE and  $\Delta{B}_{{\rm{LOS}}}$ are related, and their locations often overlap between 0-60\%. 
Not all locations coincide, thus potentially indicating differences in their origin.
We find that the WL area is related to the flare class by a power law and extend the findings of previous studies, that the WLE is related to the flare class by a power law, to also be valid for C-class flares. To compare unresolved (Sun-as-a-star) WL measurements to our data, we derive a method to calculate temperatures and areas of such data under the black-body assumption. The calculated unresolved WLE areas improve, but still differ to the resolved flaring area by about a factor of 5-10 (previously 10-20), which could be explained by various physical or instrumental causes. This method could also be applied to stellar flares to determine their temperatures and areas independently. 

\end{abstract}


\section{Introduction}

The whole electromagnetic spectrum, from UV to X-rays, is seen to increase during flares, indicating energy dissipation via radiation. 
A significant part of the total energy radiated during flares is thought to be emitted by enhancement of the continuum radiation, which is called  white-light (WL) emission in the optical wavelengths \citep[e.g.,][]{Neidig1989, Hudson2011}. The exact mechanism, however, is yet unclear and measurements have indicated hydrogen recombination continua and enhanced H$^{-}$ emission, which allow us to probe the energy dissipation heights. Large flares tend to show large continuum enhancements
\citep[e.g,][]{Hao2012A&A,Kleint2016ApJ...816...88K,Kuhar2016,Watanabe2017ApJ,Namekata2017ApJ...851...91N}, but WL emission has been observed even for small flares \citep{Jess2008}. A main open question is why certain flares show WL emission, while others do not, which may give hints on energy dissipation and particle transport.

Ground-based observations often suffer from variable seeing conditions, which may strongly affect the detection of WL emission through the varying contrast. For a statistical study of WL emission, it is, therefore, advantageous to use space-based observations with constant cadence and contrast. The Solar Dynamics Observatory is well suited to study these types of phenomena due to its continuous observations, stability, and full-disk observations taken by the Helioseismic and Magnetic Imager \cite[HMI,][]{Scherrer2012}.

Observations have shown that permanent changes of the photospheric magnetic field (hereafter \blos{}) are a common phenomenon during flares \citep{Sudol2005,Petrie2010,CastellanosDuran2018}, even during C-class flares \citep{CastellanosDuran2018,Bi2018ApJ}. The fact that they are observed at photospheric level suggests energy transfer from the coronal reconnection site via a yet unclear mechanism. Surprisingly, the so far only observation of chromospheric magnetic field changes showed the \blos{} to differ significantly from those in the photosphere \citep{Kleint2017ApJ}.
By analyzing 5 flares, a previous study found that changes in the horizontal magnetic field strength and in the mean continuum emission tend to by co-spatial and co-temporal \citep{Song2016ApJ...826..173S}.
The present work is the first statistical study about the relationship between the WL emission and the permanent changes of the photospheric magnetic field with a sample ranging from C- to X-class flares.

\section{Observations and methods} \label{sec:obs}
\subsection{Flare sample and data reduction}
We analyze 75 flares with a large energy range and a wide distribution on the solar disk to search for the connection between the WL emission and changes of the magnetic field during flares.
The same sample of flares was previously used for the study of \citet{CastellanosDuran2018} and includes 1 B-, 19 C- 37 M- and 18 X-class flares that occurred during the 24th solar cycle. The cosine of the heliocentric angle for the flare sample ranges from $\mu=0.27-0.97$. To analyze the temporal evolution of the magnetic field and the continuum emission, we use data from HMI. We analyze the HMI observables {\it hmi.M\_45s} (magnetograms) {\it hmi.Ic\_45s}  (continuum intensity) with a cadence of 45\,s and plate scale of 0\farcs504 pixel$^{-1}$.  HMI samples the spectral region around the \ion{Fe}{1} 6173.3 \AA\ absorption line at six wavelength points with a FWHM of the transmission profiles of 76 m\AA. Since the size of the host active region (AR) and the flare changes from one event to the other, we adjusted the field of view (FoV) to fully contain the AR. FoVs in our sample range from 80\arcsec$\times$80\arcsec{} to 300\arcsec$\times$300\arcsec{}.

\begin{figure}[tbh]
\includegraphics[width=.48\textwidth]{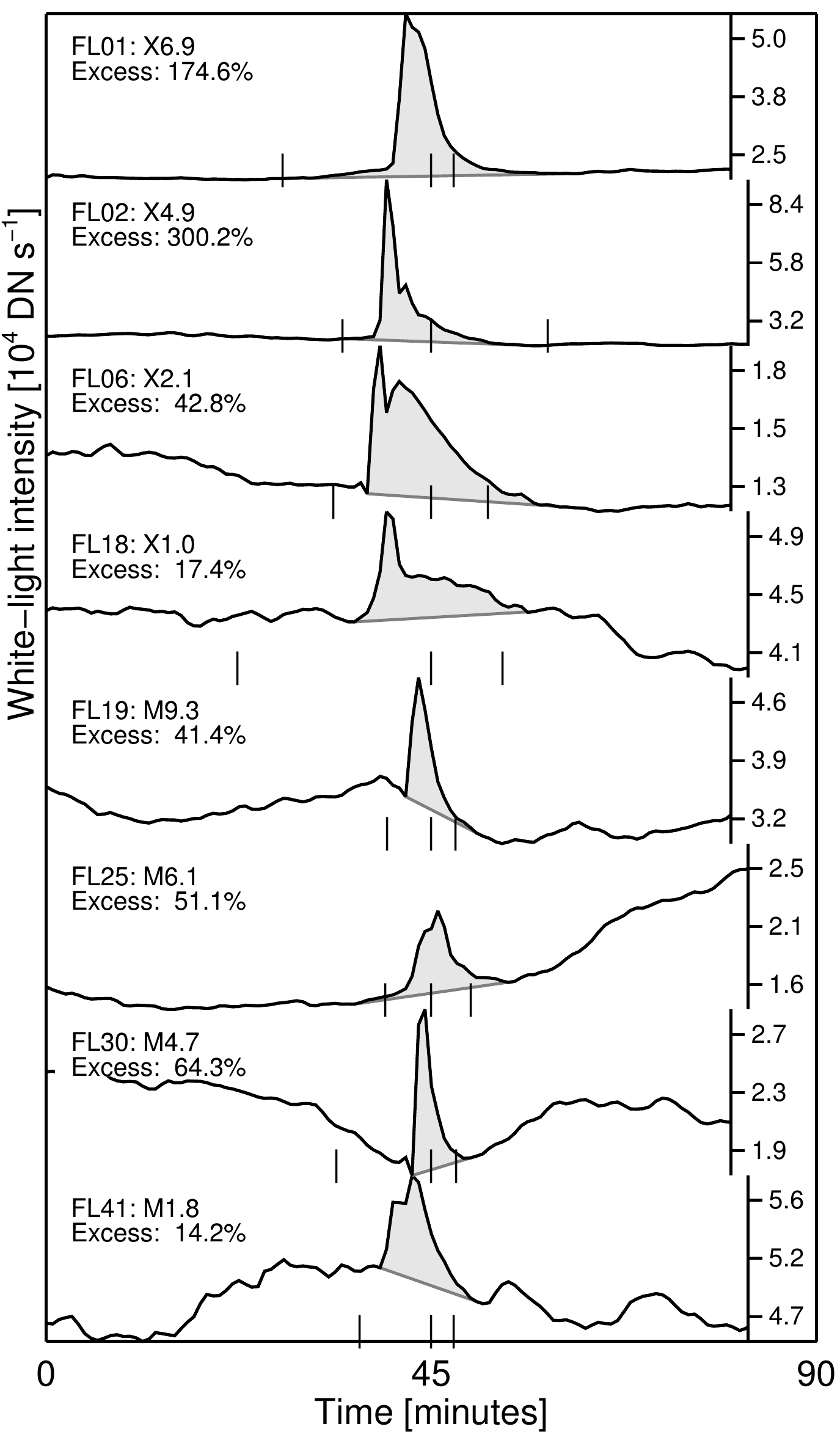}\vspace{-2mm}
\caption{Examples of HMI continuum intensity light curves with WL enhancements for 8 pixels in 8 different flares sorted by their GOES classes from X6.9 (top) to M1.8 (bottom). 
Each light curve shows the temporal evolution of a single pixel, which was selected based on the maximum enhancement. The WL enhancement is shown as the gray areas between the HMI light curve and the line that intercepts the start and end time of the WL emission for every single profile. Vertical lines indicate the GOES times of the flares. The event identifier `\texttt{FL}'\# refers to the index in Table 1 of \citet{CastellanosDuran2018}. To obtain our quantity ``WL enhancement'', we summed the grey areas of all WL pixels of a given flare.  \label{fig:wle}}
\end{figure}

The continuum images were corrected for limb darkening to second order. For a given pixel with coordinates $(x_i,y_j)$ with respect to the solar disk center $(x_c,y_c)$, the corrected intensity is given by
\begin{equation}
I_{ij}^{\rm{corr}}=I_{ij}^{non-corr}/C_{ij},
\end{equation}
where $C_{ij}$ is the limb darkening function
\begin{equation}
    C_{ij} = 1-u_{\lambda}-v_{\lambda}+
    u_{\lambda}\cos(\Theta)+
    v_{\lambda}\cos(\Theta)^2,
\end{equation}
where $\Theta=\sin^{-1}(\sqrt{(x_i-x_c)^2+(y_j-y_c)^2}/R_{\odot})$. $u_{\lambda}$ and $v_{\lambda}$ are wavelength dependent parameters, which at $\lambda=6173.3$ \AA{} are equal to $u_{6173.3}=0.836$ and $v_{6173.3}=-0.204$ \citep{Allen1976}. 

To transform the units of the HMI intensity images from DN s$^{-1}$ to physical units \ergs{}, we use the Atlas solar spectrum at 6173\,\AA{} \citep{Neckel1994svsp}.  At this wavelength, the intensity at disk center is $0.315\times10^7$\,erg\,s$^{-1}$\,cm$^{-2}$\,sr$^{-1}$\,\AA{}$^{-1}$. When averaging ten images around the peak of the flare over an area of 100 arcsec$^2$, the SDO/HMI intensity at disk center is $60000\pm300$\,DN\,s$^{-1}$.   Therefore, the conversion factor is
\begin{equation}
\alpha={52.5}\text{\,[erg\,s}^{-1}\text{\,cm}^{-2}\text{\,sr}^{-1}\text{\,\AA{}}^{-1}]/[\text{DN\,s}^{-1}].  
\end{equation}

\subsection{Identifying WL flares}\label{pixelmethod}

There are different methods to determine whether a flare has WL emission. Our goal is to find its location and how much WL radiation is released during a flare.

One method to determine the location of the WL in recent studies \citep[e.g.,][]{Kuhar2016} uses the spatial relationship between the WL and the hard X-rays (HXR) emitted by non-thermal electrons when they deposit their energy by collisions in the lower layers of the solar atmosphere \citep[e.g.,][]{Matthews2003,Krucker2015}. However, this relationship does not imply (for the majority of the flares) that the area covered by the WL relates 1:1 with the HXR contours. In addition, this assumption depends on co-observations of the flare by two different instruments such as SDO/HMI and the Reuven Ramaty High Energy Solar Spectroscopic Imager \citep[RHESSI,][]{Lin2002}.

We used three different methods to first identify if a flare is a WL flare (WLF). Method 1 is based on the running difference between two consecutive intensity images. In method 2, we took an image close to the GOES start time and subtracted it from all intensity images. This method has also been used previously, but its drawback is that the WL signal in some cases can be hidden depending on the self-evolution of the AR. Method 3 consisted of running differences, but with the adaptation that the subtracted image is either the preceding or subsequent image, depending on maximum WL emission during the duration of the flare $\pm$5 minutes before the start and end times reported by GOES. Method 3 is meant to avoid artifacts which arise from local maxima and minima of the light curves. For a frame $j$ before the maximum of the WL emission, the running difference is $I(j+1)-I(j)$. For a frame after the WL emission maximum, the running difference is $I(j)-I(j+1)$. 

Flares with strong WL emission are usually detected by all three methods. But we considered a flare to be a WLF if any of the methods gave a positive result.
For the flares where the WL emission was weak, we visually checked all light curves that have enhancements larger than $1\sigma$ in the differences for any of the three methods. $\sigma$ is calculated for each pixel individually during the pre-flare phase between the start of the observing window and the GOES start time ($\sim30$\,min on average). Flares with clear strong enhancements, or enhancements that passed the visual inspection test, were labeled as WLFs.

Once a WLF was identified, we applied a stricter criterion to reliably identify the pixels and thus the area that showed WL emission.
After different tests, we concluded that only those pixels whose running difference maximum exceeded 4$\sigma$ of the light curve can be selected as WL pixels, a criterion, which also holds for weak WLFs (see Figures~\ref{fig:int_kernelchanges_wlf}-\ref{fig:int_kernelchanges_wlf2}).

Figure \ref{fig:wle} shows examples of light curves occurring during eight flares that range from X6.9 (top) to M1.8 (bottom). The WL enhancement is shown by the gray area and the excess is given for each pixel. Vertical lines represent the GOES start, maximum, and end times. 

\subsection{ Flare emission}
To quantify the WL we used two different proxies. The WL excess  accounts for how bright the emission is ($I_{{\rm {WL}\,max}}$) in comparison with the pre-flaring conditions ($I_{{\rm pre}}$), and the WL enhancement is a proxy for how much WL is radiated during the flare (=the area under the curve minus the background; see grey area in Figure~\ref{fig:wle}). We calculated both quantities for each WL pixel for the limb-darkening corrected light curve from HMI.

The excess per pixel is given by
\begin{equation}\label{eq:excess}
{\rm {WL\,excess }}=\frac{I_{{\rm {WL}\,max}}-I_{{\rm pre}}}{I_{{\rm pre}}}.
\end{equation}

\begin{figure}[!tbh]
\includegraphics[width=.48\textwidth]{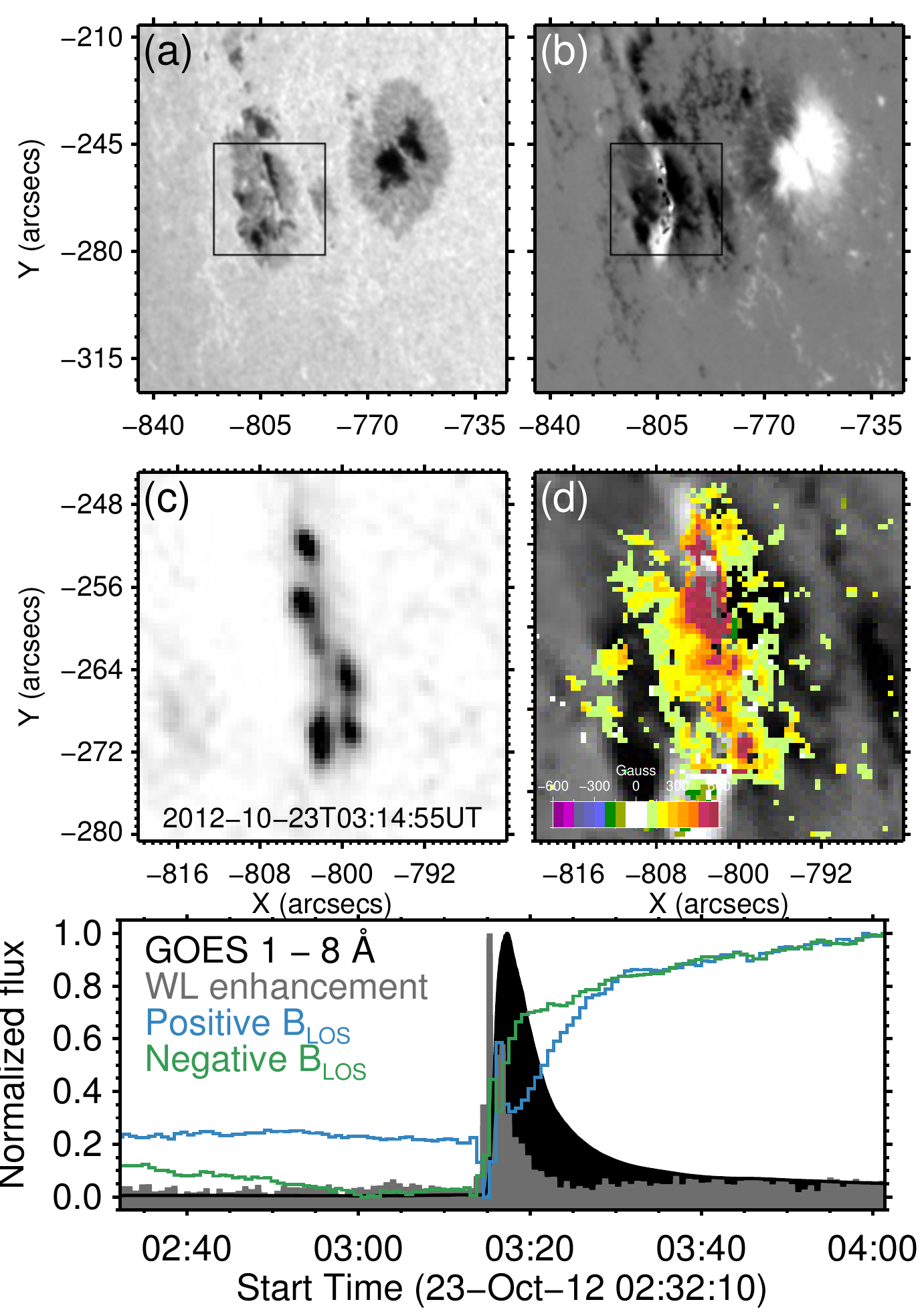}\vspace{-2mm}
\caption{Summary of the analyzed observables. Top panels show the HMI continuum image (a), and the magnetogram (b). The squares denote the FoV shown in the middle panels. The difference continuum image is presented in panel c, and panel d shows the location of \blos{} overplotted on the magnetogram. The bottom panel shows the temporal evolution of the positive (blue), and negative (green) line of sight magnetic field, as well as the total WL enhancement (gray) of the flare and the GOES soft X-ray light curve (black). All quantities in the bottom panel are integrated over the entire FoV.   \label{fig:changes}}
\end{figure} 

To calculate the WL enhancement, we first need to estimate the background. Instead of assuming a constant background, which is not a good assumption in most of the cases (see Figure~\ref{fig:wle}), we interpolated linearly between the start and end of the WL emission. We did not constrain the sign of the background's slope, which can be positive or negative (see Figure~\ref{fig:wle}). The background evolution was calculated for each pixel independently since the time of WL emission can vary between pixels during the same event. The WL enhancement was determined as the area under the curve that covers the WL emission ($I_{\rm WL})$ minus the background ($I_{\rm {backg.}})$, i.e.

\begin{equation}\label{eq:excess}
{\rm WL\,enhancement}=\sum_{\rm WL \, pixels}\sum_{t}I_{\rm WL}(t)-I_{\rm backg.}(t),
\end{equation}
where $t$ varies for each pixel depending when the pixel displays WL emission.

We are aware that the continuum intensity observed by HMI is a reconstruction using an MDI-like algorithm \citep{Couvidat2016SoPh} that is based on a simple approximation and that might be inaccurate for strong flares \citep[e.g.,][]{Svanda2018ApJ} due to varying spectral line profiles that are not seen with HMI's spectral resolution. Therefore, the uncertainty of the intensity due to these line-profile changes during flares cannot be determined without knowing the fully resolved spectral profile. For this reason, we avoid giving error bars that most likely are unrealistic \citep[see for example,][]{MartnezOliveros2014SoPh-transients,Svanda2018ApJ,Sadykov2020ApJ}.  However, we believe that the location and areas of the enhancements are relatively accurate, since this quantity is binary.

\begin{figure*}[htbp]
\begin{center}
\includegraphics[width=1.\textwidth]{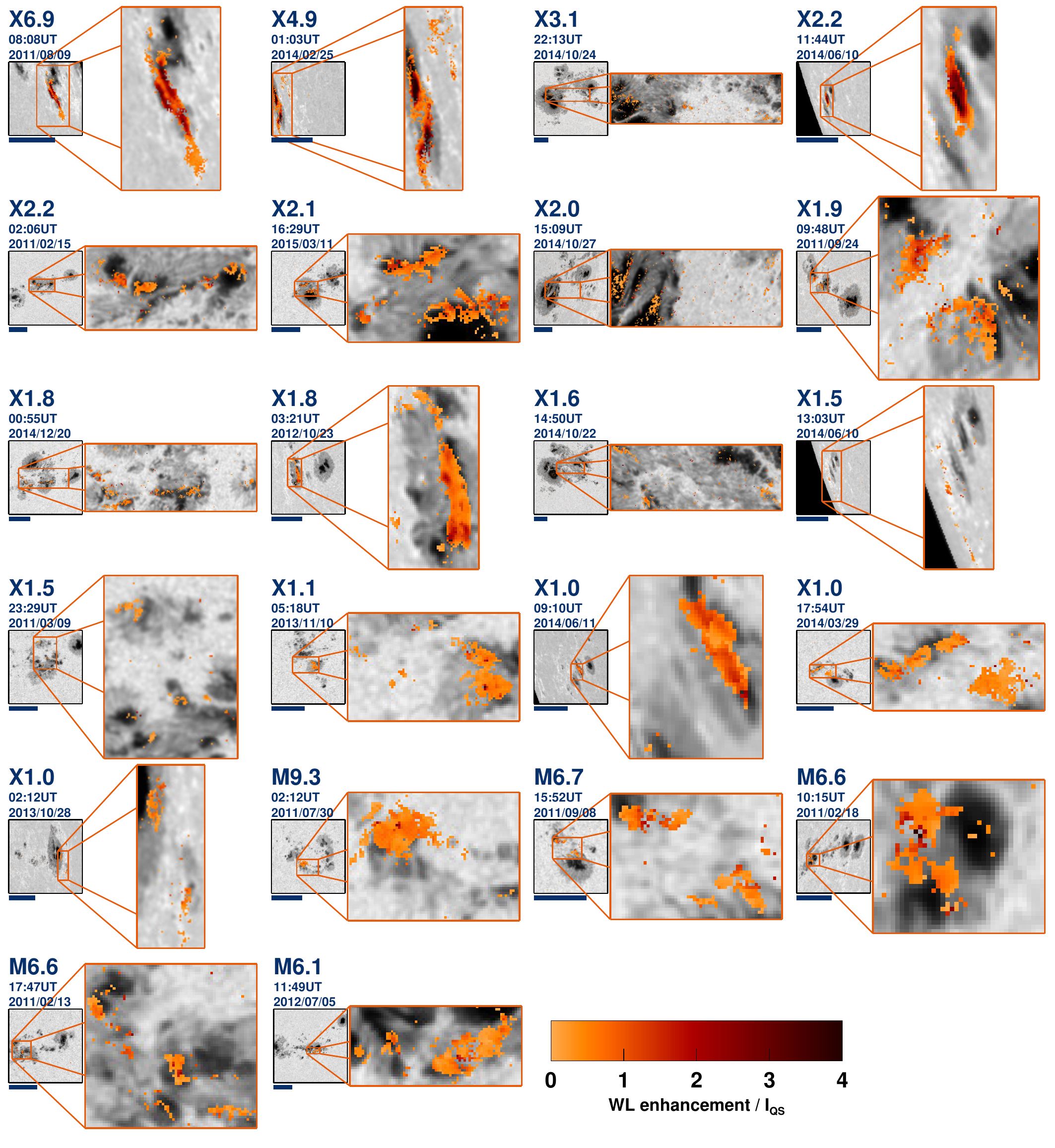} 
\caption{Flares in our sample that showed WL emission and \blos{}. The WL enhancement is saturated at 4 times the quiet Sun intensity at disk center because of the high dynamic range from C- to X-flares, following the color bar on the bottom right. 
Scaled images are shown by the blue bar with a length of 50\arcsec{}. The average of the quiet Sun intensity was calculated for each flare at disk center during the pre-flare phase over an area of $\sim$(10\arcsec{})$^2$.  Figure~\ref{fig:int_kernelchanges_wlf_mag}  shows a comparison of WL and \blos\ locations of the same events.\label{fig:int_kernelchanges_wlf} }
\end{center}
\end{figure*}

\begin{figure*}[htbp]
\begin{center}
\includegraphics[width=1.\textwidth]{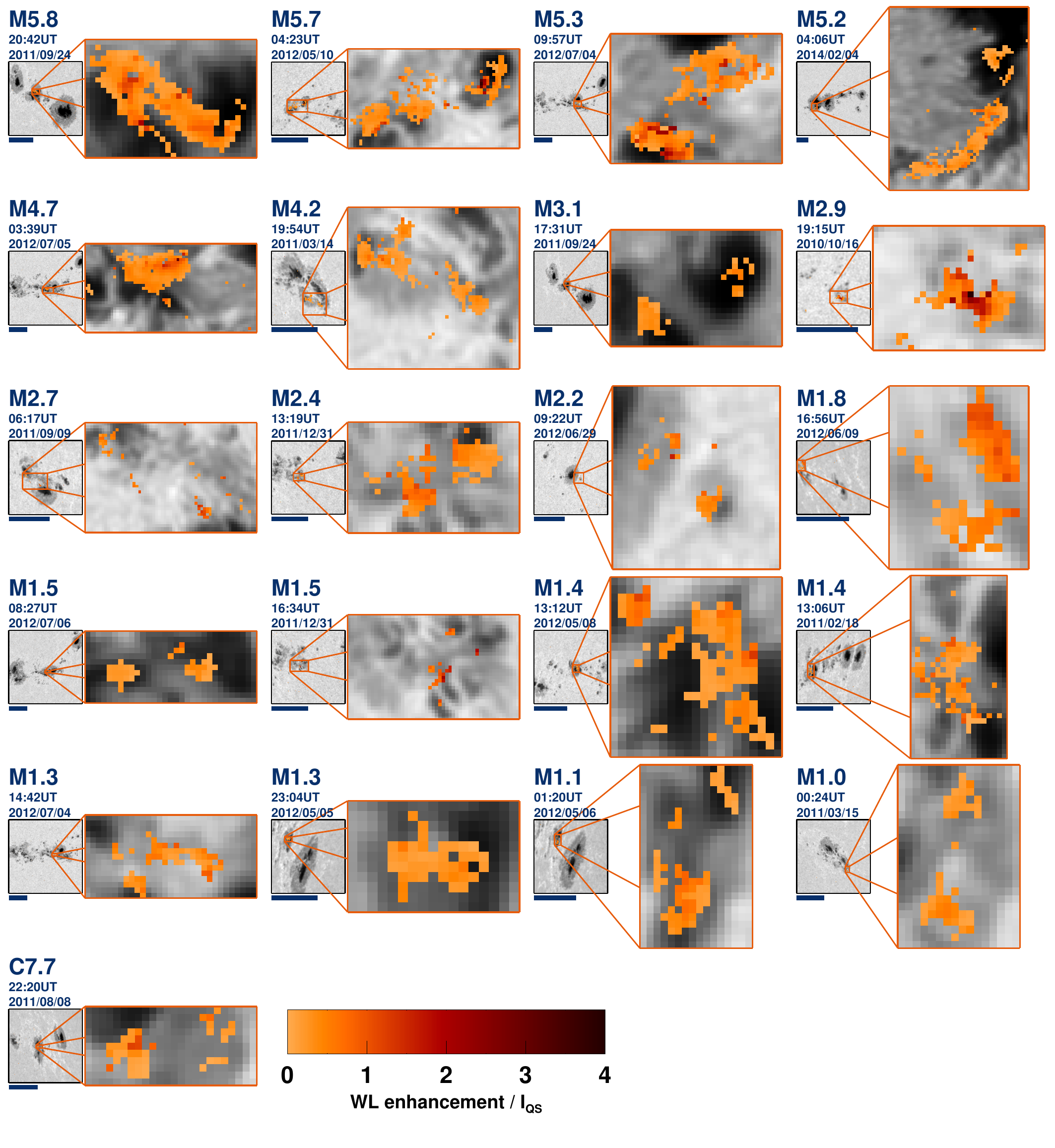} 
\caption{Same layout as Figure~\ref{fig:int_kernelchanges_wlf}.\label{fig:int_kernelchanges_wlf2} 
See also Figure~\ref{fig:int_kernelchanges_wlf_mag2}  for a comparison with the magnetic field change locations. \label{fig:int_kernelchanges_wlf_2}}
\end{center}
\end{figure*}

\subsection{Magnetic field changes}
We use the magnetic field changes determined by \citet{CastellanosDuran2018} to compare the locations of magnetic field changes and WL emission. The authors used the following method to derive the magnetic field changes. They fit the temporal series of $B_{\rm LOS}$(t)  measurements within the FoV with a stepwise function \bt{}$=B_l(t)+B_{step}(t)$ for each pixel, as given by \cite{Sudol2005}. 
This function is divided into two components: a linear part that takes into account the background field
\begin{equation}
B_l(t)=a+bt,
\end{equation}
and the stepwise part that accounts for a possible stepwise change 
\begin{equation}
B_{step}(t)=c\left\{1+\frac{2}{\pi}\tan^{-1}\left[n(t-t_0)\right]\right\},
\end{equation}where $2c$, $n^{-1}$ and $t_0$ describe the size, duration, and midpoint of the step, respectively. Light curves with changes smaller than 80 G or which do not show clear irreversible changes are excluded. Figure \ref{fig:changes} shows an example of the data analyzed in this work during the X1.8 flare (\texttt{SOL2012-10-23T03:17}). The continuum and magnetograms are displayed in \ref{fig:changes}a-\ref{fig:changes}b. The WL enhancement and the location of \blos{} are shown in panels \ref{fig:changes}c-\ref{fig:changes}d. The bottom panels show the temporal evolution of the soft X-ray (black), line-of-sight magnetic field (blue and green), and the WL emission (gray).  For a full description on the conditions and methods to characterize the \blos{}, please refer to \citet{CastellanosDuran2018}.

\section{Results}\label{sec:result}

\subsection{WL occurrence}
During 44 of 75 events (59\%), WL enhancements were observed. Figures \ref{fig:int_kernelchanges_wlf} and \ref{fig:int_kernelchanges_wlf2} shows the AR and a zoom of the locations of the WL emission during the 43 events that showed both WL emission and \blos{}, overplotted on the AR that flared. The approximate WL enhancements per flare reach up to $\sim2\times10^7$\,\ergs{}.
The WL kernels tend to be elongated and have the highest emission in the middle that diffuses towards the edges.  Figures \ref{fig:int_kernelchanges_wlf_mag} and \ref{fig:int_kernelchanges_wlf_mag2} show a comparison of WL and \blos{} of the same events. The overlapping areas are shown in red.
There is one C5.6 flare (\texttt{SOL2011-05-27T16:43}) that showed clear WL emission, although \blos{} could not be located. The opposite case is also observed in an X-class flare 
(\texttt{SOL2012-07-12T16:50}), 
where HMI did not detect WL emission.

\subsection{Dependence of WL emission on $\mu$}

Due to the limb darkening, the intensity near the limb is lower than the intensity at disk center. \citet{Kuhar2016} reported a decrease in the WL flux for flares near the limb ($>$900\arcsec$\approx72^{\circ}$). In our analysis, where limb-darkening was removed from all images, we do not find a relationship between the WL emission and $\mu$-angle of the flare. It is worth noting that the longitude of the majority (70/75) of the flares in our sample is smaller than $72^{\circ}$, which might explain this result. The only flare in our sample with longitude close to $90^{\circ}$ is the C5.6 flare that is accompanied by WL emission but does not show \blos{}. The limb darkening removal may not be entirely accurate when considering potential chromospheric WL emission, but the fraction of chromospheric WL emission to photospheric WL emission is not yet well known and the limb darkening removal is the currently best approximation.
\begin{figure}[htbp]
\begin{center}
\includegraphics[width=.48\textwidth]{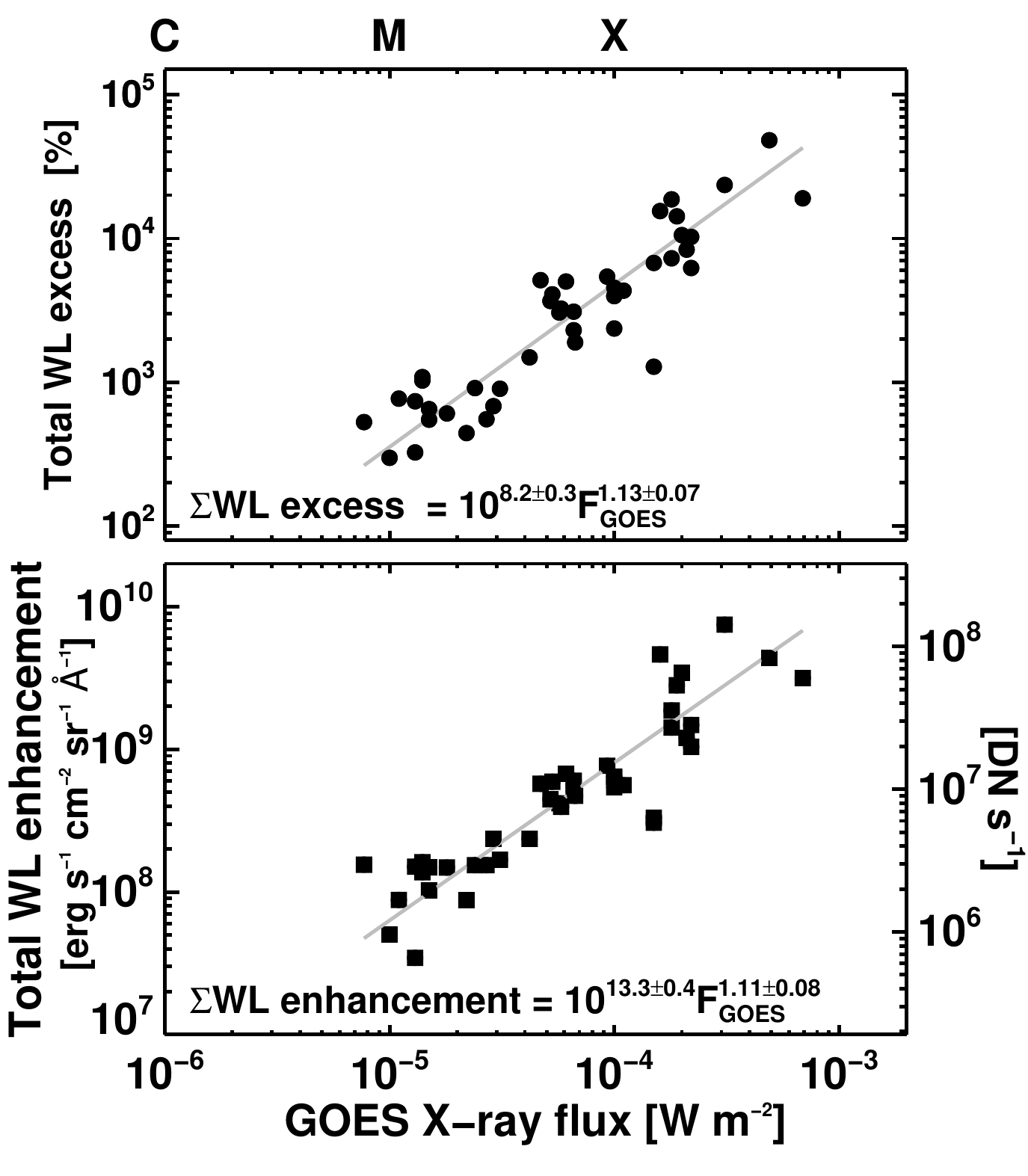}\vspace{-3mm}
\caption{The total WL excess (top) and WL enhancement (bottom) for each flare as a function of the GOES X-ray flux.  Straight lines are the best-fit of a power-law model ${\rm F}_{{\rm WL}}\propto  {{\rm F}}_{{\rm GOES}}^{\delta}$. 
\label{fig:wlstatistics} 
}
\end{center}
\end{figure} 

\subsection{Dependence of total WL emission on flare energy}

More energetic flares have been found to display stronger WL emission. 
\citep[][]{Kahler1982JGR,Neidig1983SoPh,Matthews2003,Wang2009RAA,Kuhar2016,Namekata2017ApJ...851...91N}.
Figure \ref{fig:wlstatistics} shows the total WL excess and enhancement as a function of the GOES X-ray flux (${{\rm F}}_{{\rm GOES}}$).  ``Total'' means the sum over all pixels that showed WL emission during each flare.
The best fit of a power-law distribution is given by  

\begin{align}\label{eq:wlflux}
\Sigma{\rm WL}({{\rm Excess}}) &= 10^{8.2\pm0.3}\ {\rm F}^{1.13\pm0.07}_{{\rm GOES}},\\
\Sigma{\rm WL}({{\rm Enhancement}}) &= 10^{13.3\pm0.4}\ {\rm F}^{1.11\pm0.08}_{{\rm GOES}},
\end{align}

where we used the bisector regression method to calculate the best fit and the error bars \citep{Isobe1990ApJ}. Despite the exponent for the total WL excess being slightly larger compared to that of the total WL enhancement, both fits follow the same behavior. Considering that the WL emission covers large areas (Figure~\ref{fig:wle}), we opted for the sum instead of reporting just the pixel with maximum intensity, or the average of all WL pixels, which might have skewed the interpretation of the phenomena. The Total WL excess strongly depends on the location of the source (see discussion in \S4.7) and therefore is not fully representative of the strength of the WL enhancement. For stars, where the radiation is integrated over the entire stellar disk, the measured excess is our proxy for the actual energy input due to WL.

\begin{figure}[tbph]
\includegraphics[width=.48\textwidth]{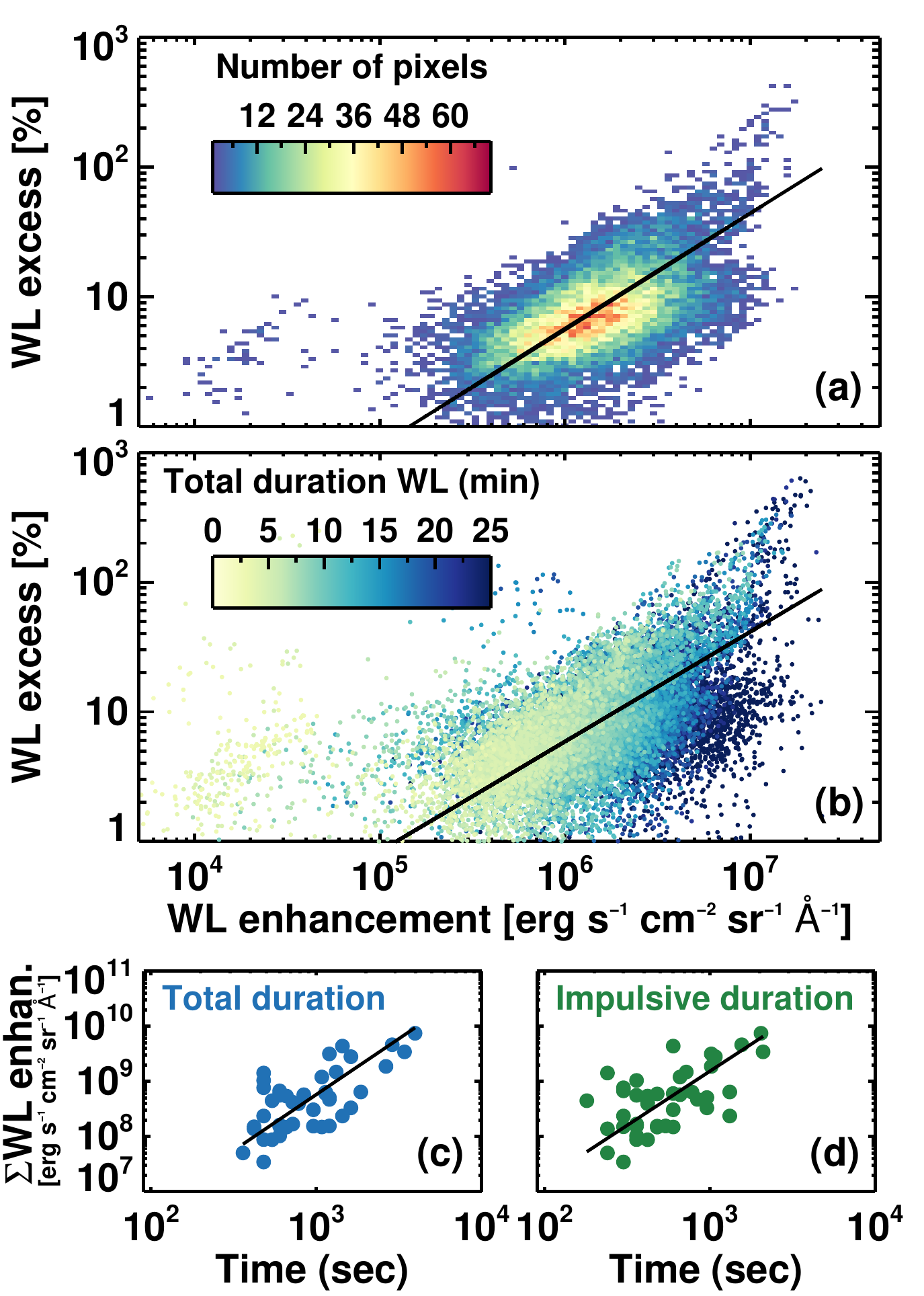}
\vspace{-4mm}
\caption{Scatter plots of the comparison of the WL excess and the WL enhancement for all pixels where we detected WL emission. Colors in panel (a) represent the number of pixels, as given by the color bar. The black line is the best fit of $\log y=(-1.92\pm0.03)+(0.89\pm0.01)\log(x)$. One should be careful when interpreting the WL emission by the WL excess since these quantities are not linearly related. Colors in panel (b) represent the total duration in minutes of the WL emission per pixel. The bottom panels show the total WL enhancement as a function of the total duration of the flare (panel c) and the duration of the impulsive phase (panel d) as estimated by using the reported GOES times for each flare. 
  \label{fig:excess2enhancement}}
\end{figure}

To investigate the difference between the excess and the enhancement for all flares in the sample, we use a scatter plot for these quantities in Figure~\ref{fig:excess2enhancement}. The colors in panel a) represent the number of pixels at each bin and the black line displays the fit to the data given by $\log y=(-1.92\pm0.03)+(0.89\pm0.01)\log(x)$. The non-linear relationship shows that one should be careful when comparing the excess and the enhancement to describe the WL emission during flares. While the excess describes how strong the emission is (when comparing with the pre-flare conditions) and where this emission occurs (see discussion below), the enhancement accounts for the total WL emission over time that can span several minutes (see Figure~\ref{fig:wle}).

In 43/44 of the WLFs, we could obtain the full GOES 1-8 \AA{} light curve of the flare. We observed in all WLFs that the WL emission peaks before the GOES maximum. We then calculated the derivative of the GOES light curve, as an approximation of the HXR emission via the Neupert effect \citep{Neupert1968ApJ}. In 93\%  (40/43) of the flares, the peak of the GOES derivative is within $\pm$1 frame of the observed WL peak emission. Because HMI data products are available for every 45 seconds, our temporal resolution element is 90 seconds, and therefore we cannot constrain the time better than $\pm$1 HMI frames with respect to the time observed by GOES.

Based on Figure~\ref{fig:excess2enhancement}a there seems to be an upper limit on how much a pixel radiates in WL: few times $10^7$\,\,erg\,s$^{-1}$\,cm$^{-2}$\,sr$^{-1}$\,\AA{}$^{-1}$. To test if it is related to the duration of the WL emission, we color-coded in panel (b) the duration of the WL occurrence in the flare. The WL enhancement depends on the total duration of the emission, i.e., the longer the WL emission, the larger the WL enhancement in general. On the other hand, the WL excess does not depend on the duration of the emission and does not seem to be related to how fast the energy is radiated during the flare. This suggests that the WL enhancement carries more information when compared with the WL excess since it combines both, temporal evolution and strength of the emission. Also, the upper limit may be related to sample bias, since just 6 flares in our sample have a duration longer than half an hour. Long duration events  are observed less frequently  on the Sun \citep[][]{Shibata1996AdSpR}.

In addition, the excess indirectly depends on the cadence of the observations as pointed out by \citet{Wang2009RAA}, since the likelihood to observe the flare at the exact time when the WL emission is at the maximum reduces with lower cadence. The excess, or in some studies also known as contrast, is a relative measurement. Different studies have used different quantities as the base-line intensity. Some studies used the average over the WL region \citep[e.g.,][]{Matthews2003}, or the neighboring quiet Sun \citep[e.g.,][]{Hudson2006SoPh,Jess2008}, or the intensity at the same location in non-flaring conditions \citep[e.g.,][]{Wang2009RAA,Kuhar2016}, which make  a direct comparison of the excess between different studies difficult.

\begin{figure}[tpb]
\begin{center}
\includegraphics[width=.47\textwidth]{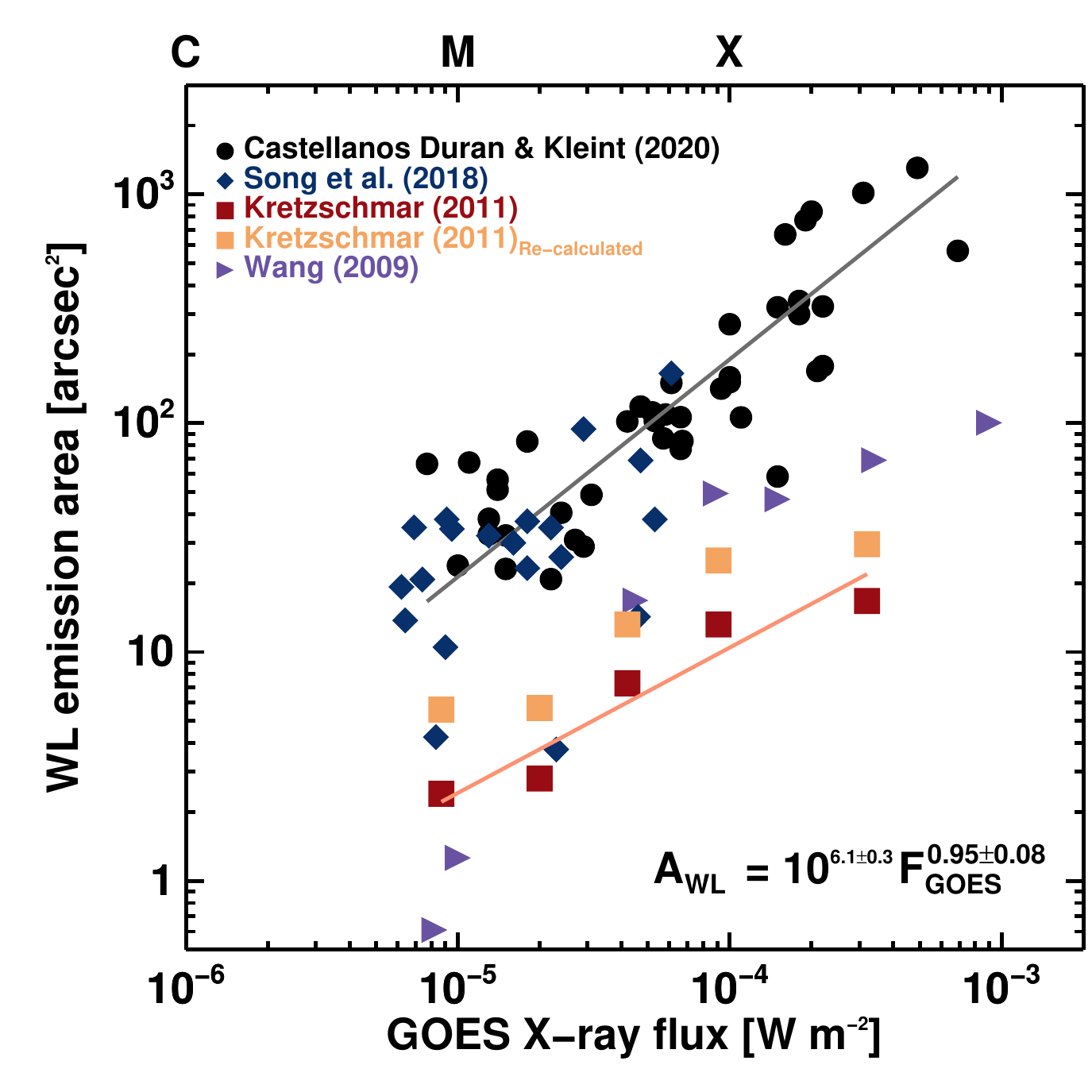} \vspace{-7mm}
\caption{Area of the WL emission as function of the GOES X-ray flux (black circles). 
Straight lines are the best-fit of a power-law model ${\rm A}_{{\rm WL}}\propto  {{\rm F}}_{{\rm GOES}}^{\delta}$. Blue diamonds represent WL areas reported by \protect\citet{Song2018AA613A69S} and the purple triagles are the areas estimated by \protect\citet{Wang2009RAA}. Note that \citet{Wang2009RAA} reported the equivalent area, which is expected to be smaller than the total area. Red squares show the WL emission areas estimated with Sun-as-a-star observations measured by \protect\citet{Kretzschmar2011}, and brown squares are the \citet{Kretzschmar2011} re-calculated values using the method derived in the appendix \ref{sec:appendix_temp}. The recalculation places the Sun-as-a-star values closer to the spatially-resolved values.\label{fig:areachanges}}
\end{center}
\end{figure}

\subsection{Area covered by WL}
The total area covered by the WL emission (${\rm A}_{{\rm WL}}$) for the flare sample ranges from 6.3 Mm$^2$ to 822.2 Mm$^2$. The area was calculated based on the size of an HMI pixel of $0\farcs5 \times 0\farcs5$ and by summing the number of pixels that exhibited WL emission at any time during the flare. Figure \ref{fig:areachanges} reveals that the area of the WL emission is related with the GOES X-ray flux  following a power-law, given by 

\begin{equation}\label{eq:areas}
{\rm A}_{{\rm WL}} = 10^{6.1\pm0.3}\ {\rm F}^{0.95\pm0.08}_{{\rm GOES}}. 
\end{equation}

This relationship does not depend on the location of the flare after correcting the areas for foreshortening. 

We devised a method to estimate the WL area and temperature based on the luminosity enhancement, which is described in Appendix A and we apply it to the sun-as-a-star WL enhancements reported by \citet{Kretzschmar2011}. The resulting non-spatially resolved WL area is slightly below our measured WL areas from HMI as shown in Figure~\ref{fig:areachanges}.

\subsection{Correlation of WL and \blos{}}
\begin{figure}[tbp]
\begin{center}
\includegraphics[width=.47\textwidth]{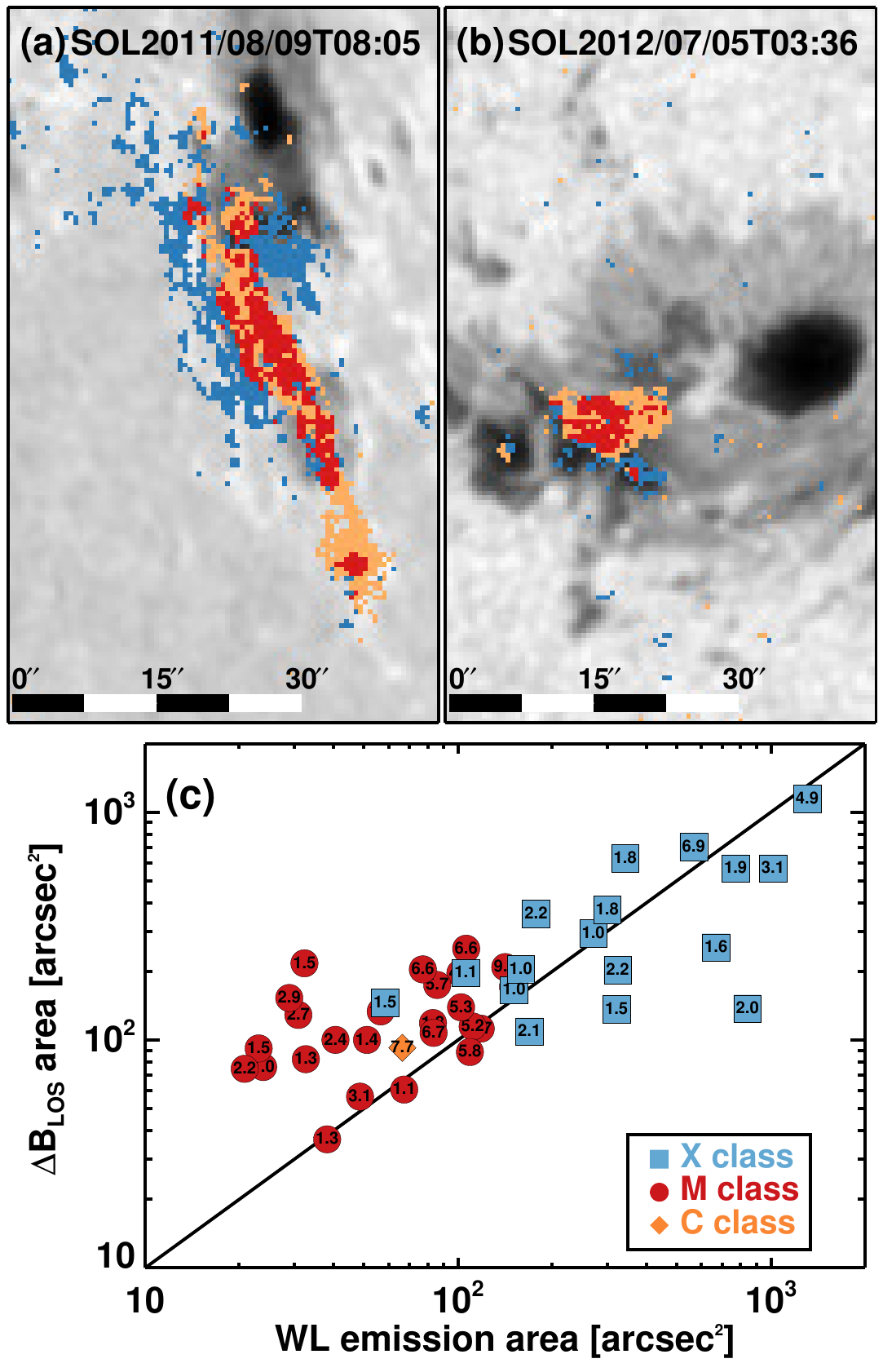}\vspace{-2mm}
\caption{Spatial relationship between the area of the WL emission and \blos{}. The top panels display the location of \blos{} (blue) and WL emission (orange) during flares X6.9 - \texttt{SOL2011-08-09T08:05} (a) and M4.7 - \texttt{SOL2012-07-05T03:36} (b). Red pixels denote positions where both \blos{} and WL emission coincided. Panel (c) shows the comparison between the total area of WL emission and \blos{}. The black line is not a fit, but a visual help for the 1:1 area relation. The detection limit is 0.25 arcsec$^2$, given by HMI's pixel size, and for our sample the area covered by the \blos{} tended to be larger than the area of the  WL emission. \label{fig:wl2bchangeslocation}}
\end{center}
\end{figure}

\begin{figure}[htbp]
\begin{center}
\includegraphics[width=.49\textwidth]{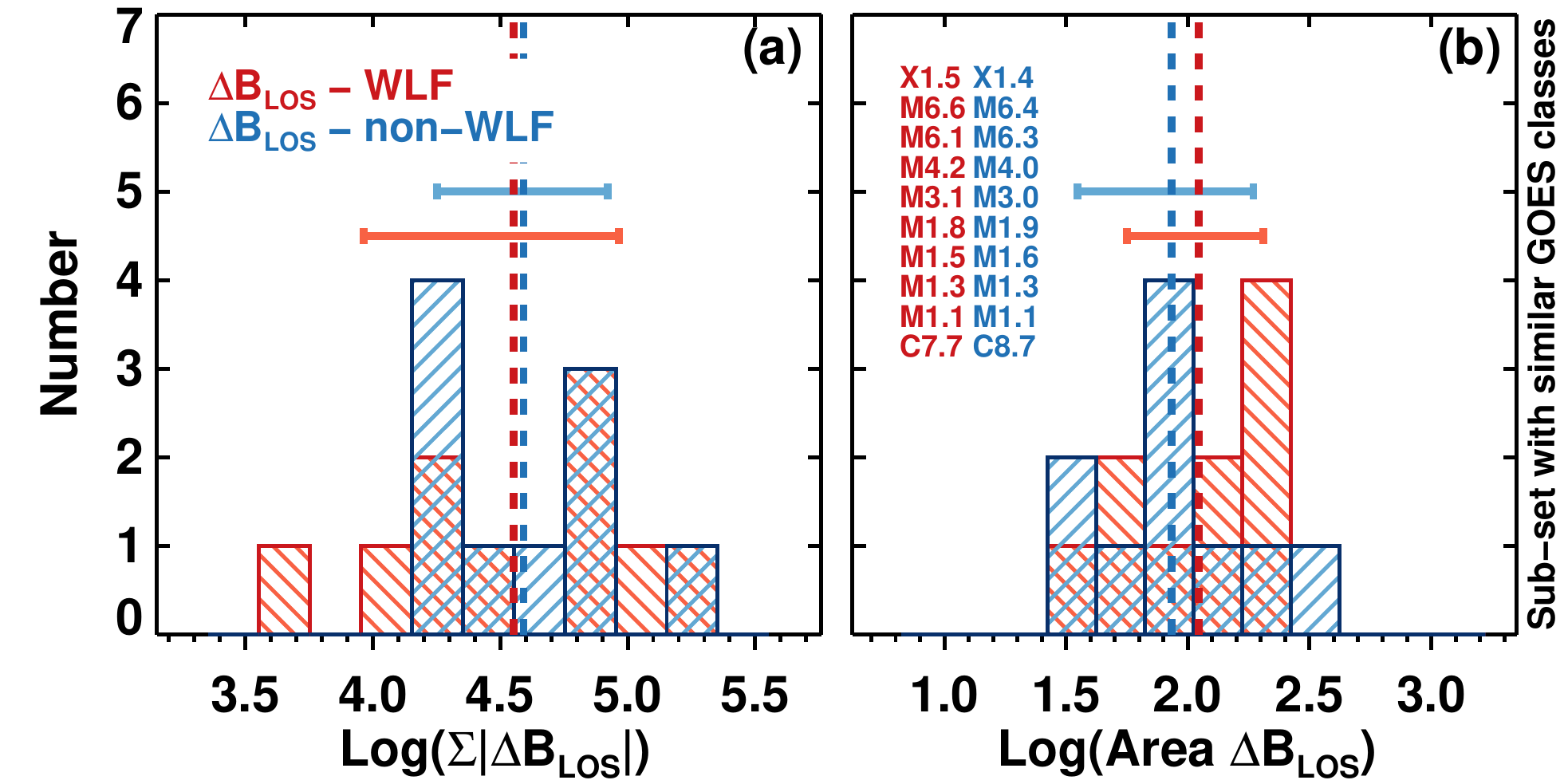} 
\caption{The total unsigned \blos{} (left) and area of \blos{} (right)
for an unbiased sample in terms of GOES class of WL flares (red) and non-WL flares (blue) that showed magnetic field changes. The dashed lines indicate the medians. The horizontal bars show the $50^{\rm th}\pm34^{\rm th}$ percentile to approximate $\pm1\sigma$. We do not find any difference between WLF and non-WLF.
\label{fig:hirtononwlf}}

\end{center}
\end{figure}

\begin{figure*}[htbp]
\begin{center}
\includegraphics[width=1.\textwidth]{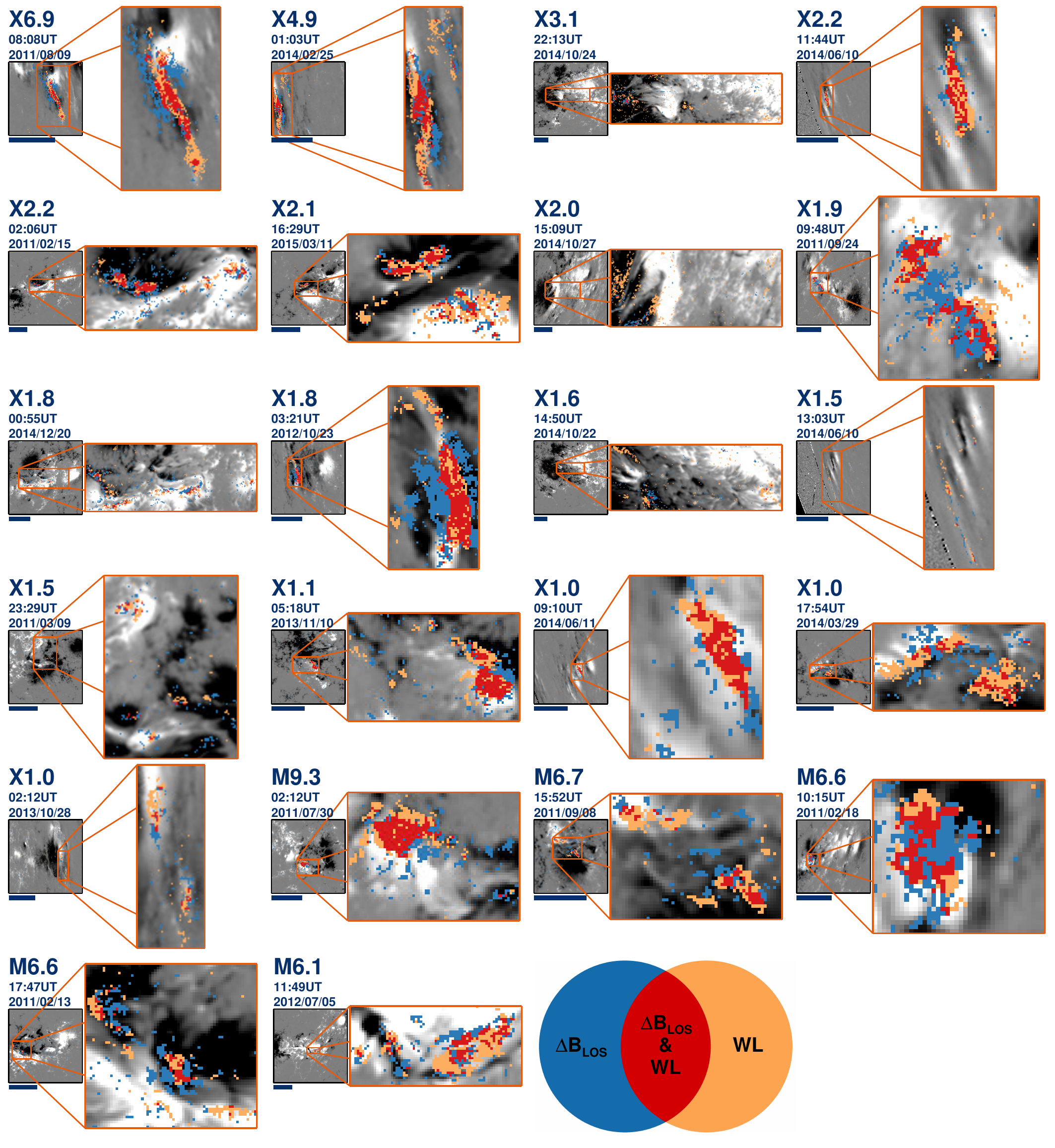} 
\caption{Same layout as Figure~\ref{fig:int_kernelchanges_wlf2}, but the background image is replaced by the line-of-sight magnetic of the same region clipped at $\pm800$\,G. The coloured pixels show the location of \blos{} (blue), WL emission (orange) and the overlapping areas (red). \label{fig:int_kernelchanges_wlf_mag}}
\end{center}
\end{figure*}

\begin{figure*}[htbp]
\begin{center}
\includegraphics[width=1.\textwidth]{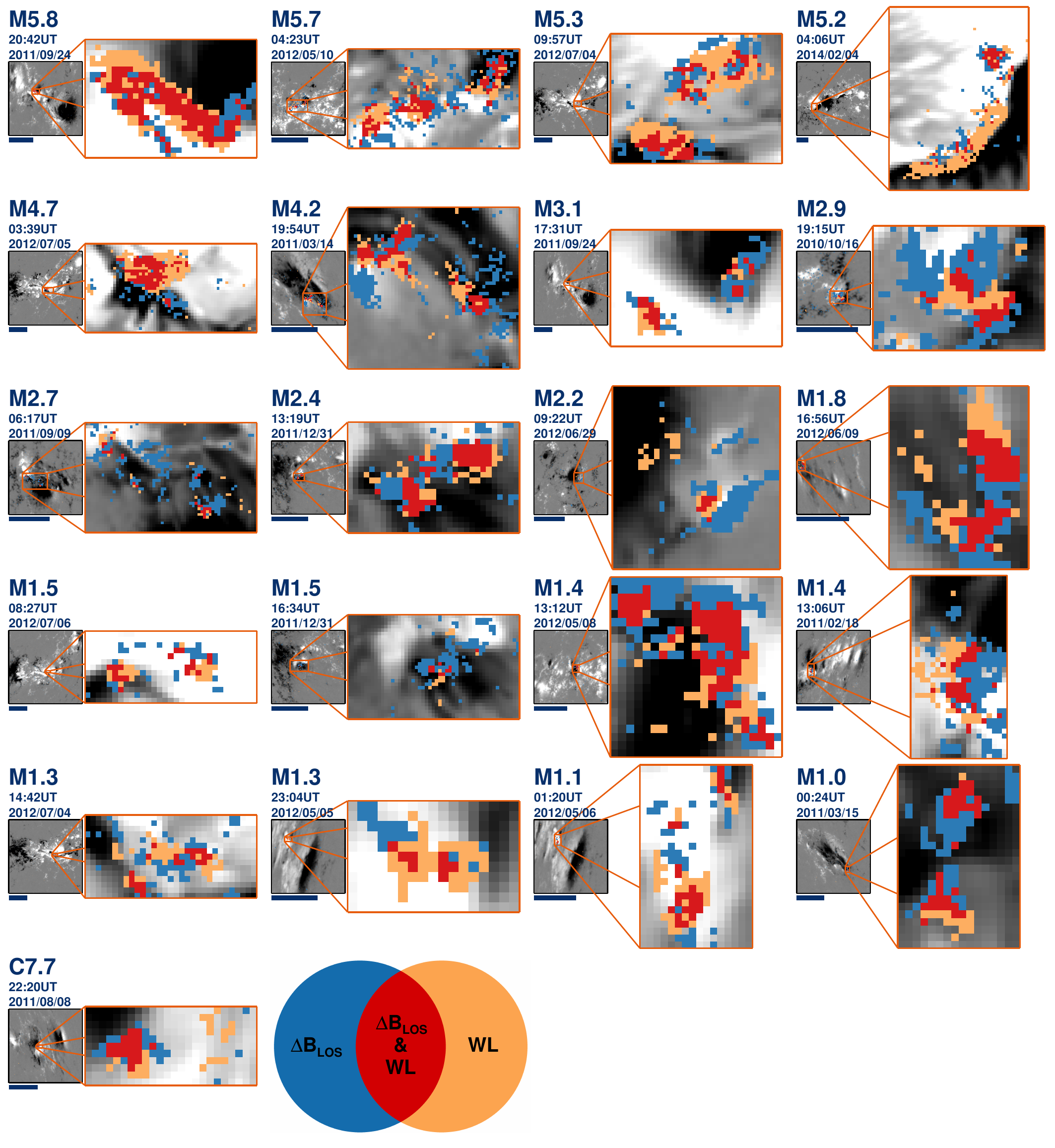} 
\caption{Same layout as Figure~\ref{fig:int_kernelchanges_wlf_mag}. The coloured pixels show the location of \blos{} (blue), WL emission (orange) and the overlapping areas (red). \label{fig:int_kernelchanges_wlf_mag2}}
\end{center}
\end{figure*}

It is an open question which processes cause WL emission and \blos{} and if they are related. We therefore investigate their spatial relation. 43 of 44 flares that showed WL emission were also accompanied by \blos{}. WL emission and \blos{} are usually observed near each other and often overlap. The images on top in Figure~\ref{fig:wl2bchangeslocation} display two examples of the location of \blos{} and WL emission during X6.9 (a) and M4.7 (b) flares. Color-coded pixels show the location of the WL emission (orange) and \blos{} (blue). Locations, where WL emission and \blos{} coincide, are pictured in red.  If both \blos{} and WL emission are observed, larger \blos{} are found near, or at the same places of WL emission. Figure \ref{fig:wl2bchangeslocation}c compares the total area covered by \blos{} and the WL. The GOES classes of the flares are represented by the colors and numbers inside the symbols.   31/43 (72\%) of the events show that ${\rm A}_{\BLOS{}}$ are larger than ${\rm A}_{{\rm WL}}$. We also investigated if the overlap area is related to the flare strength (GOES class) or the heliocentric angle $\mu$, but there was no correlation. With one exception, the WL and \blos{} areas overlap between 0\% (for the C5.6 flare without \blos{}) and 65\% within our analysis, which contains \blos{}$>80$ G and WL pixels with a 4$\sigma$ maximum. The mean overlap area for the entire sample of flares is $(29.2\pm16.4)\%$ and $(20.8\pm14.4)\%$ of the WL area and \blos{} area, respectively. 

We investigated if there is a difference between the flares that showed both WL emission and \blos{}, with respect to the events without WL emission. We selected an unbiased sample of 20 flares with \blos{}, 10 with WL emission, 10 without WL emission, always pairs with about a similar GOES classes. We calculated the distribution of unsigned flux and the \blos{} areas for both sets, as shown in Figure~\ref{fig:hirtononwlf}. There seems to be no difference in the medians between WLF and non-WLF. Because of the small sample size and thus not necessarily a normal distribution, we approximated $\sigma$ by taking the $50^{\rm th}\pm34^{\rm th}$ percentiles of the cumulative distribution. It is very important to select such a balanced sample of flares, because when performing the same analysis on the entire flare sample, which is biased because strong flares generally show stronger WL, and larger areas of WL and \blos{}, one may wrongly conclude that there is a difference between WLF and non-WLF. It may be difficult to define a balanced sample, because stronger flares tend to show more WL emission and therefore our sample included mostly M-flares.

\section{Discussion}

\subsection{WL Occurrence}
In our study 94.4\% (17/18) of the X-class flares are associated with WL emission, 64.9\% (24/37) of the M-class flares, 10.5\% (2/19) of the C-class flares. 
The only exception for strong flares is the X1.4 flare (\texttt{SOL2012-07-12T16:50}), where no WL emission was found. X-class flares without WL emission have been previously observed \citep[e.g.,][]{Watanabe2017ApJ}. For flares with GOES classes larger than M8/M5 28/26 of 29 flares showed WL emission. This result does not fully agree with previous findings \citep{Matthews2003, Kuhar2016}, who found that all flares larger than M8/M5 are accompanied by WL emission. However, the total duration of the analyzed X1.4 flare is 1:53\,hours and the time difference between the GOES start and peak times is 1:12\,hours. This may be an indication of how slow the energy is injected into the lower atmosphere during long-duration flares, and thus, may be important when observing WL emission, as pointed out by \citet{Kuhar2016} and \citet{Watanabe2017ApJ}. For this reason, \citet{Kuhar2016} explained that they focused their analysis on flares with short and intense hard X-ray emission.

When analyzing the relationship between circular ribbons flares and the WL emission, \citet{Song2018ApJ...867..159S}  observed WL emission in 100\% (10/10), 61.8\% (21/34), 8.5\% (4/47), and 0\% (0/1) for flares with GOES classes of X, M, C, and B, respectively.  These percentages are consistent with the results presented in the current work.  Table \ref{tab:wlf} in the appendix \ref{sec:occurrencewlf} summarizes the appearance of WLFs between different studies. From the low-number statistics, it is found that X-, M-, and C-class flares are accompanied by WL emission in 83.7\%, 51.2\% and 11.0\%, repectively.

 53\% of all flares in this study showed WL emission, which is almost twice as frequent as the 28\% previously reported by \citet{Song2018AA613A69S}. The discrepancy is due to a sample bias between the two studies in terms of the energy of the flares. \citet{Song2018AA613A69S}'s sample does not contain any X-class flares, and 12/23 (52\%) M-class and 8/47 (17\%) C-class flares show WL emission, which is consistent with our results.  But why do we observe WL more often in large flares than in smaller ones? \citet{Jess2008} suggested that the reason could be due to the spatial resolution of the telescope and its sensitivity. The power-laws found for the WL excess/enhancement and the WL area, strongly indicate that the source of the {\it missing} WL emission in some flares could be a purely observational problem.

\subsection{WL Timing}
The WL emission starts after the GOES start time and in all observed flares reaches its maximum before the GOES peak, in agreement with previous observations \citep[see Figure~1 of][]{Kuhar2016}. The times of the WL emission of each pixel are not directly correlated with the GOES times (see Figure\,\ref{fig:wle}), probably because the particle acceleration, which seems to be related to WL emission \citep[e.g.,][]{Fletcher2007ApJ,Krucker2015}, occurs at different times in different locations, while the GOES times are based on an integration over the whole Sun. 
Previous studies have shown that about half of the flares \citep{veronigetal2002} tend to follow the so-called Neupert effect \citep{Neupert1968ApJ}, an observed correlation between the SXR radiation and the time-integral of the HXR radiation. If the mechanism for WL emission and HXR is closely related, as suggested by single observations \citep[e.g.][]{krucker2011}, and if the Neupert effect is valid, one should expect a close relation in timing between the derivative of the GOES SXR (as an approximation of the HXR) and the WL emission. Our sample shows that in 93\% of the cases (40/43 analyzed WLFs), this relation is indeed valid with the caveat that our temporal resolution of the WL data is only 90 seconds. This analysis would benefit from having faster cadence WL data and carrying out the comparison directly with the HXR emission, e.g. from RHESSI, instead of using a proxy, whose reliability may only be valid in half of the cases. While we do not find any contradiction for the close relationship of WL and HXR, further studies are required to analyze their detailed relationship. 

\citet{Watanabe2017ApJ} statistically found that shorter flares tend to have stronger WL emission, suggesting the impulsivity is important. However, panels (c) and (d) in Figure~\ref{fig:excess2enhancement} show that the total WL enhancement relates with the total duration of the flare and the duration of the impulsive phase. At least in our flare sample, this correlation is in fact opposite to the behavior found by \citet{Watanabe2017ApJ}.

\subsection{WL structure}
The WL emission shows a similar shape in all flares: bright kernels and decreasing intensity towards the edges of the usually elongated WL emission (see Figure~\ref{fig:int_kernelchanges_wlf}). The fine structure observed in the WLF agrees with previous findings \citep{Matthews2003,Hudson2006SoPh}.

\citet{Kuhar2016} used the assumption that WL kernels are well correlated with the RHESSI footpoints, based on past studies, which have shown very good agreement in single flares \citep{Krucker2015}, but a statistical study has never been performed. We plan to perform a future study, which may clarify the relation between particle acceleration and WL.

\begin{figure}[!tb]
\begin{center}
\includegraphics[width=.47\textwidth]{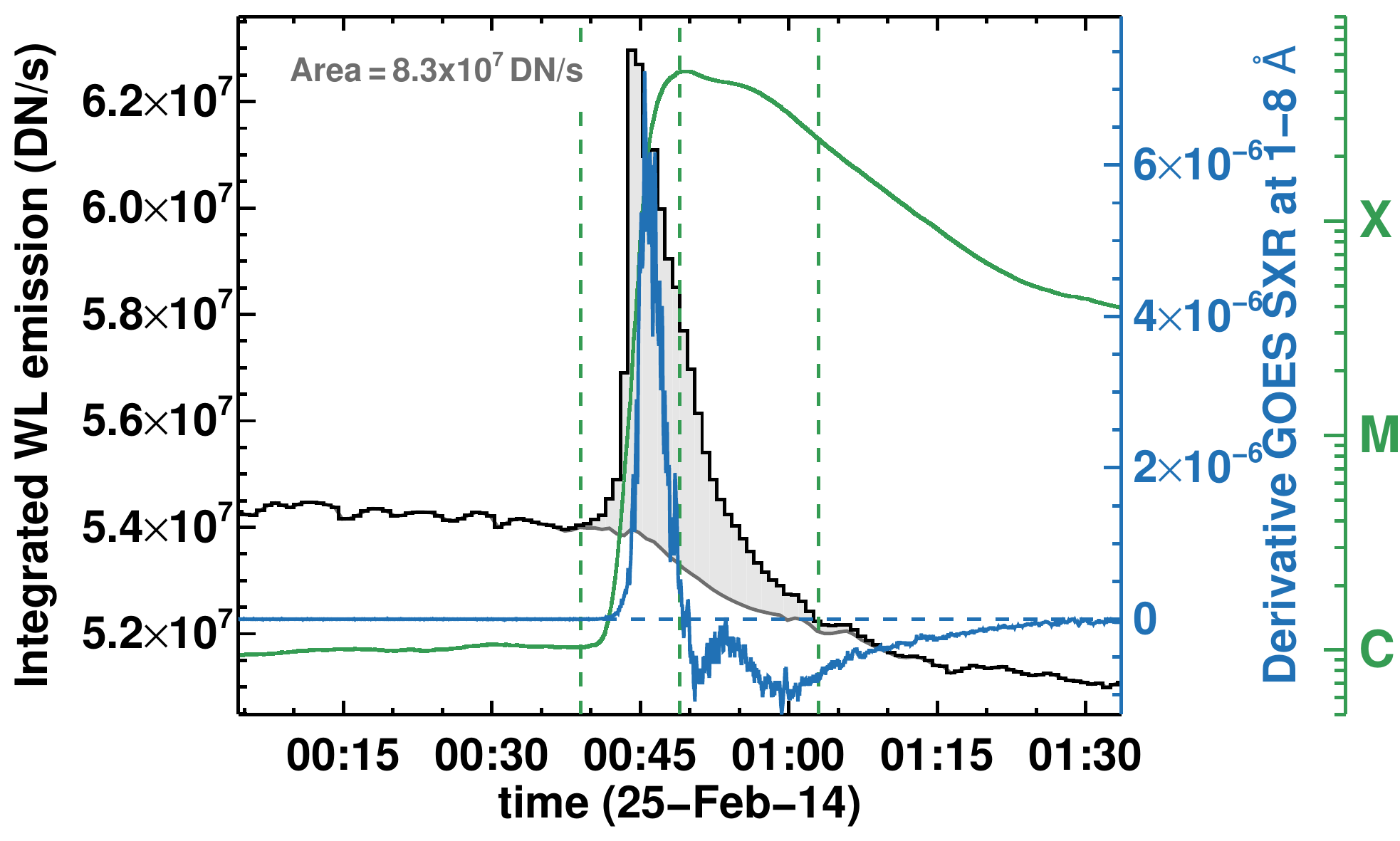} 
\caption{The integrated WL emission of the solar flare  \texttt{SOL2014/02/25T00:49:00} (black line). We summed over all pixels that showed WL emission in Figure~\ref{fig:int_kernelchanges_wlf}.  The dark-grey line is the time dependent background and the light-grey area is the total WL emission of the flare. The green lines are the GOES light curve and the vertical dashed lines mark the GOES start, peak and end of the flare. The blue line is the derivative of the GOES light curve. This example should illustrate our method to determine the WL emission in comparison with other methods, which often take the peak value minus the pre-flare value without a temporal integration.  \label{fig:Neupert}} 
\end{center}
\end{figure}

\subsection{WL and GOES Soft X-rays}

We found that the total WL enhancement integrated in time and the GOES class are related following a power-law with a power index $\delta_{{\rm CDK20}}=1.11\pm0.08$ (see Figure~\ref{fig:wlstatistics}).

To compare this with different authors is difficult because of the different methods. While we integrate over time and thus analyze the total WL enhancement, \citet{Kuhar2016} and \citet{Namekata2017ApJ...851...91N} looked at the WL emission inside a given RHESSI contour, but only at one time step  (what we call WL excess). 
In our study, the duration of the emission (=the integration time) varies for each pixel and only WL pixels were taken into account, which were identified as described in Sect.~\ref{pixelmethod} and not based on any contours. Both \citet{Namekata2017ApJ...851...91N} and our study approximated the evolution of the background flux, but with different methods.
In Figure~\ref{fig:Neupert} we demonstrate how our method accounts for the time-variation of the WL emission and the background.
The black line in Figure \ref{fig:Neupert} displays the total continuum emission inside the pixels shown in Figure~\ref{fig:int_kernelchanges_wlf} for the X4.9 flare \texttt{SOL2014/02/25T00:49:00}. The dark-gray line is the total background and varies in time because the number of WL pixel and the noise changes in time. 
The gray area between the continuum emission and the background is what we report as the total WL enhancement (integrated in time). The green line in Figure~\ref{fig:Neupert} denotes the GOES 1--8 \AA{} light curve and the vertical lines mark the GOES times, while the blue line is the derivative of the GOES light curve. Because the 
 WL emission appears at different locations at different times  (see for example Figures~4 of \citet{Kerr2014}, or of \citet{Sharykin2017ApJ})
 we considered it more accurate to quantify the WL emission as an integrated quantity in time. Nevertheless, \citet{Kuhar2016}\footnote{We estimated the correlation for the flares presented by \citet{Kuhar2016} using the values in their Table~1 and applied a multiplication by 2.8 due to a missing scaling factor after private communication with the authors.} and \citet{Namekata2017ApJ...851...91N} both also found similar power-laws for the relationship between their definition of WL flux and the soft X-ray emission.

While we find a good correlation between the total WL emission integrated in time, and the total WL area as a function of the GOES class, \citet{Toriumi2017ApJ} did not find a good correlation between ribbon areas and GOES class. They derived the ribbon areas based on SDO/AIA 1600 \AA\ data and determined a correlation coefficient of $r^2_{\text{T}17}=0.23$ between the ribbon area and the GOES class for 51 flares (see their table 2). The authors suggested that it might be related to sample bias since they analyzed flares $\geq$M5-class, but our power-laws also hold for sub-sets of strong flares. It may be possible that the ribbon area is more related to the atmospheric heating evolution, than the actual flare strength (as determined by GOES class). Because soft X-ray emission traces the thermal evolution, one may expect a correlation of the ribbon area with the integrated GOES emission of a given flare, but the relation of WL and flare ribbon areas is out of the scope of this work.

\subsection{Location of the WL and \blos{}}

The co-spatial relationship between the WL and the \blos{} found in this study is consistent with other analyses of smaller flare samples \citep[e.g.,][]{Zhang1994ApJ,Kleint2017ApJ,Sun2017ApJ,Song2016ApJ...826..173S} and now statistically verified. We found that the WL emission and the \blos{} are related and often overlap. But because the overlap is not 1:1 and its percentage varies in the different flares, there may be a different origin causing the WLE and \blos{}. 
The areas are of the same order of magnitude, and there does not seem to be a difference in \blos{} areas or flux between WLF and non-WLF. Therefore, we conclude that the WL and the \blos{} are linked in some way and their origin might differ. However, the mechanisms causing WL and \blos{} are still unclear, which makes an interpretation difficult.

\subsection{Possible sources of error and reliability of HMI's data during flares}

The HMI continuum intensity measurements are well known to produce artifacts when fast phenomena occur, such as flares \citep[e.g.,][]{MartnezOliveros2014SoPh-transients}. Different processes such as mass motions, gradients in temperature and the magnetic field and in some cases lines in emission might produce complex signatures in the spectral lines \citep[e.g.,][]{Mauas1990ApJS,Zuccarello2020ApJ}. For example, \citet{Svanda2018ApJ} studied the intensity of the 6 HMI passbands and the derived continuum emission during a large flare and compared it to Hinode/SP spectra and simulated HMI spectra where the iron lines were in emission. They concluded that HMI data are adequate to trace the evolution of the flare ribbons but might have severe shortcomings in estimating the continuum intensity, due to the simplistic MDI-like algorithm that HMI uses in determining the \texttt{Ic.45s} data-products. HMI samples the \ion{Fe}{1} line at only six different wavelength points, at different times, and without a filter located at true continuum. Even though this is not ideal for flare continuum measurements, HMI currently is the only instrument with continuous full-disk coverage with sufficient spatial and temporal resolution for statistical studies of the continuum emission.

In the future, the Polarimetric and Helioseismic Imager \citep[PHI,][]{Solanki2019arXiv}  will fly onboard the Solar Orbiter (SO). SO/PHI will measure the same \ion{Fe}{1} line at $\lambda$6173.3 at SDO/HMI at six different wavelength points. One of its differences to HMI is that SO/PHI has a real continuum filter at 400\,m\AA{} from the \ion{Fe}{1} line core. This corresponds to a Doppler velocity of 19 km/s from the line core, which is three times larger than the typical sound speed where the \ion{Fe}{1} line forms, which is more favorable to study the continuum enhancement during flares.

\subsection{On the bias of the excess for dark regions}\label{sec:excessbais}

\begin{figure}[thp]
\begin{center}
\includegraphics[width=.48\textwidth]{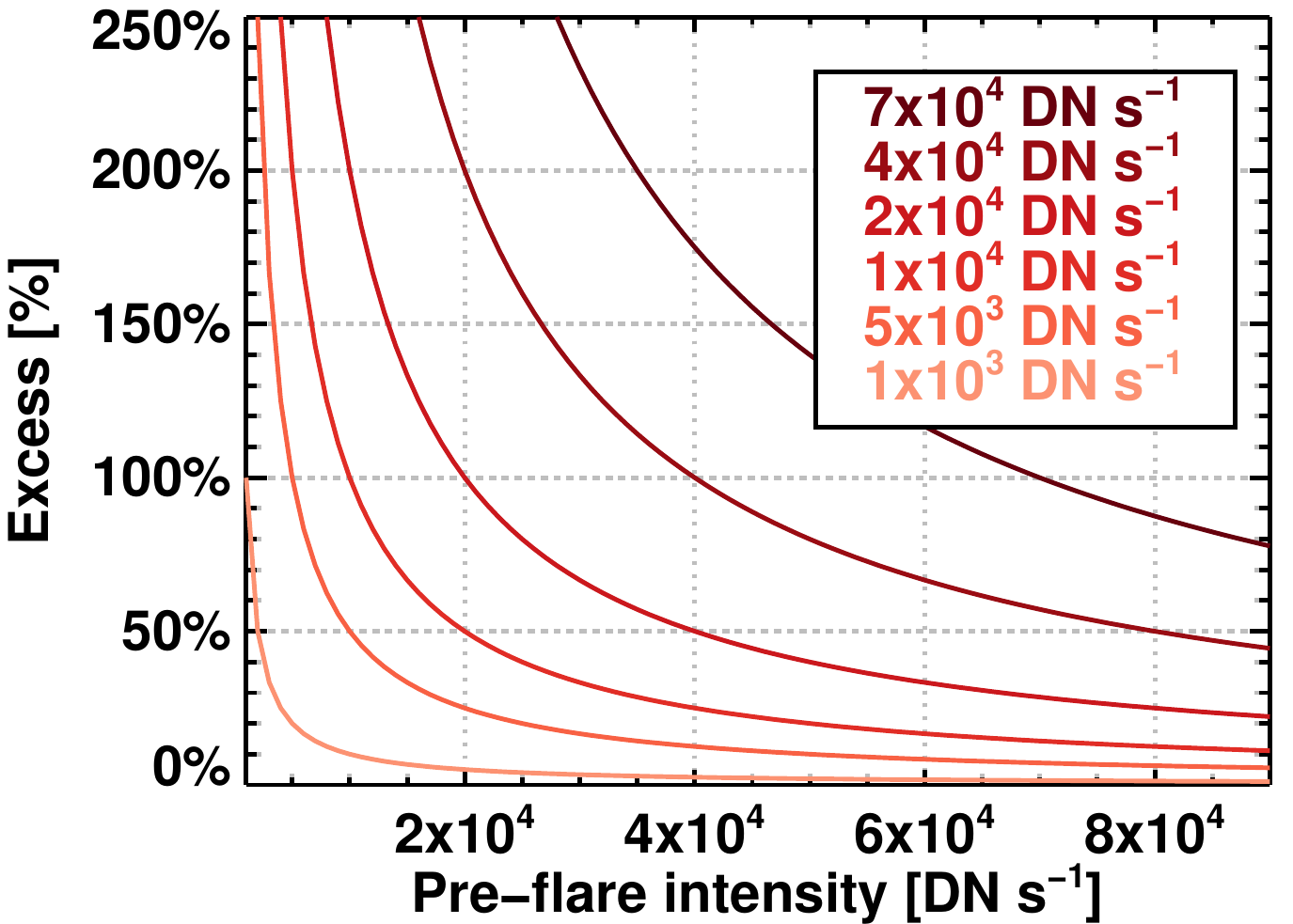} \vspace{-4mm}
\caption{WL excess that is observed for a given pre-flare intensity (on x-axis) and an assumed intensity enhancement during the flare (colored lines and legend). For example, a flare intensity enhancement of $2\times10^4$ DN s$^{-1}$ corresponds to an excess of $\sim$100-200\% in the umbra (with pre-flare intensity $1-2 \times 10^4$ DN/s), but only to $\sim33\%$ in the quiet Sun (with pre-flare intensity $6 \times 10^4$ DN/s).
\label{fig:excess2locality}}
\end{center}
\end{figure}

The excess compares the pre-flare intensity and the emission coming from the flare (Eq. \ref{eq:excess}) and this quantity highly depends on the  location of the emission. WL enhancements are seen better in umbrae and therefore there is a bias in the detection. We illustrate this in Figure~\ref{fig:excess2locality}. For a given pre-flare intensity, ranging from the umbra to quiet Sun on the x-axis, we display the observed excess for different flare intensity enhancements (colored lines), given in the legend. For a constant energy input, it is therefore easier to detect an enhancement above darker solar regions. This also explains the behavior of Figure~\ref{fig:excess2enhancement} because the WL enhancement does not depend on where the WL is produced, while the excess does.

\subsection{Resolved vs.\,unresolved WL emission areas}

The WL emission areas found in our study are consistent with previous studies of WL flares, which showed that the WL emission area increases with the energy of the flare \citep[e.g.,][]{Wang2009RAA}. In the case of stellar flares, since we cannot directly resolve the flare area, indirect approximations have to be used \citep[e.g.,][]{Maehara2012, Shibayama2013, Candelaresi2014}.
Sun-as-a-star observations have estimated the area of the WL emission using the total radiated energy  observed as the total solar irradiance (TSI) over the time that the flare occurred \citep[][hereafter K2011]{Kretzschmar2011}. In Figure~\ref{fig:areachanges} the black circles plot the WL emission areas reported by K2011. The WL emission areas measured using TSI data also follow a power-law trend ${\rm A}_{{\rm K2011}}=10^{4.11}{\rm F}_{{\rm GOES}}^{0.76}$  \citep{Kretzschmar2011, Shibayama2013}. However, the WL areas obtained by K2011 are one order of magnitude underestimated when compared with measurements using spatially-resolved solar data. 
In appendix \ref{sec:appendix_temp}, we present a refined method to calculate the flare's temperature ($T_{\text{f}}$) and area ($A_{\text{f}}$) without imposing any restriction on $A_{\text{f}}$. We re-calculated the temperatures  and areas of the flares using the values reported by K2011 with our method.

We used K2011's reported temperatures and interpolated the ratio of the Planck function at 4020\,\AA{} and 5000\,\AA{} at these temperatures to derive the observed ratios. This was more precise than interpolating them from their Figure~5. These ratios were then used as input for Eq.~\ref{eqa4}. Here we use a solar temperature of 5960\,K, which corresponds to the ratio of the two quiet Sun fluxes given in K2011 (1.67/1.83 W/m$^2$). A solar temperature of 5778\,K was also tried, but the results do not differ much from 5960\,K. Using this input for Eq.~\ref{eqa4} we derived the recalculated temperatures $T_{\text{f}}$ shown in Table~\ref{tab:kretz2011}. 

According to Eq.~\ref{eqA5}, deriving the areas requires knowing the luminosity during the flare, which is not explicitly given in K2011's paper. We reverse-engineered their observed enhancements by using their formula and their derived areas and obtained $f_b\times10^{-5}$\,W/m$^2$ for the luminosity enhancements in the blue filter.  By using their pre-flare luminosity of 1.67 W/m$^2$, our recalculated temperatures, and Eq.~\ref{eqA5}, we derive the areas shown in Table~\ref{tab:kretz2011}, which increased by about a factor of 2 compared to K2011. The re-calculated flare temperatures are $\sim$1000\,K lower. These temperatures are similar to those reported by \citet{Kowalski2017ApJ}, but we do not assume a fixed emission area (see their Appendix B).

\begin{table}[htpb]
\centering
\begin{tabular}{cccccc}
\toprule[1.5pt]
GOES & $f_b\times10^{-5}$& $T_{\text{f}}$ & $T_{\text{f}}-T_{\text{K2011}}$& $A_{\text{f}}$& $A_{\text{K2011}}$\\
class& (W/m$^2$)&(K) &(K) &(\arcsec)$^2$&(\arcsec)$^2$\\
\midrule[1.5pt]
X3.2&7.7&8304& -1041& 29.6 & 16.7\\
M9.1&5.8&7918& -1075& 25.1 &  13.2\\
M4.2&3.2&8198& -1046& 13.2& 7.3\\
M2.0&1.1&7867& -1074& 5.7& 2.8\\
C8.7&0.8&7547& -1108& 5.6& 2.4\\
\bottomrule[1.5pt]
\end{tabular}
\caption{Re-calculated temperatures and flare's areas from \citet{Kretzschmar2011}. Notice that the re-calculated temperatures are $\sim$1000\,K lower and the areas are larger and closer to the
the spatially-resolved SDO/HMI observations.\label{tab:kretz2011}}
\end{table}

Brown squares in Figure~\ref{fig:areachanges} show that the re-calculated sun-as-a-star values 
differ by about a factor of 5 from the HMI areas. There are different possibilities for explanations: 1) It is possible that the emission in the blue filter includes the contribution from spectral lines that go into emission during flares, which may increase the measured $F_{\text{Blue}}/F_{\text{Green}}$ ratio, which in turn would lead to an underestimation of the areas. 2) Because the low TSI signal required averaging many flares and the dependence of flare area vs. flare class is a power law, it is possible that the reported ``flare class" in K2011 may need to be adapted for the area calculation. 3) Because we are dealing with very small enhancements (10$^{-5}$ vs. 1.67 W/m$^2$), any instrumental effect, for example degradation, may influence the results. 4) Potentially optically-thin emission may influence our assumption of a pure black-body, however we believe that this effect is not dominant because there seems to be an intensity difference between chromospheric off-limb WL radiation \citep[e.g.,][]{heinzel2017ApJ...847...48H} compared to on-disk WL emission at the same wavelength in the visible \citep[e.g.][]{Kleint2016ApJ...816...88K}, 5) The integration time may matter significantly in the determination of the flare area because the flare ribbons are moving. Our WL areas may be higher than others because we counted every pixel that exhibited WL at any time. 6) The determined WL area depends on the spatial resolution of the instrument. Small-scale WL enhancements, which are below the size of a pixel, may be missed.

\section{Conclusions} \label{sec:discussion}

This paper presents a statistical study of WL and its relation to \blos{} using HMI data. The summary of our findings is
\begin{itemize} 

\item The WL emission and \blos{} are spatially linked but their origin is unclear. 57\% of our analyzed flares showed WL emission. Strong flares all show WL with the strange exception of one X-class flare that showed no WL but did show \blos{}. One C-flare showed WL, but no \blos{}. Therefore we conclude that both phenomena are often linked, but in a yet unknown manner.

\item The WL emission and \blos{} areas are comparable.

\item We found clear evidence that the GOES class and thus the strength of the flares is correlated with the total WL excess/enhancement, as well as with the areas affected by \blos{} and WL emission.  This is observational evidence that the amount of energy required to disturb the magnetic field is larger for larger changes \citep{Wang2010}.

\item WLFs, flares with associated \blos{}, and flares where neither of these two phenomena was detected do not show a fundamental distinction.
The lack of detection of WL or \blos{} in some flare seems to be an observational problem \citep{Neidig1989,Hudson2006SoPh,Jess2008,CastellanosDuran2018}. 
\item We present a revised method to estimate the temperature and the area of the flare under the blackbody assumption when non-spatially resolved data is used, which takes into account the flaring and non-flaring emission and areas from a star. By using observations in at least two wavelength bands, one can uniquely determine the flare area and temperature. After applying this method, we found that areas from sun-as-a-star observations become closer to spatially resolved areas, however there still is a difference of less than one order of a magnitude, which could be explained either by physics or by instrumental effects. This method updates the temperatures and areas derived by previous studies \citep{Kretzschmar2011,Shibayama2013}. This method could also be applied for stellar flares to determine their areas and temperatures independently.
\end{itemize}

\smallskip

The continuous observational coverage by HMI is highly advantageous for statistical flare studies. Statistics of magnetic field changes and of WL emission have been explored here, but their physical origins are still unclear. A future step will be to try to link them to accelerated particles, by analyzing RHESSI data. Additionally, an analysis of the magnetic field vector may reveal particular changes in the Lorentz force, which could drive the magnetic field changes \citep[see][]{Petrie2019ApJS}. Yet one observable is very rarely obtained during flares: magnetic field information at higher atmospheric layers, such as the chromosphere \citep[e.g.,][]{Kleint2016ApJ...816...88K}. Hopefully, future DKIST data will soon allow us to probe this yet largely unexplored layer.

\acknowledgments
J.\,S. Castellanos Dur\'an is deeply grateful with Prof. Benjamin Calvo Mozo for all his support and discussions on white light emission. We express our gratitude to Petr Heinzel for discussions on stellar irradiances and Adam Kowalski for discussions on flare areas and temperatures. We thank the referee for a very careful reading and suggestions that improved the quality of the manuscript. J.\,S. Castellanos Dur\'an was funded by the Deutscher Akademischer Austauschdienst (DAAD) and the International Max Planck Research School (IMPRS) for Solar System Science at the University of G\"ottingen, and the grant HERMES-26675, {\it J\'ovenes Investigadores 2014}, COLCIENCIAS - 645, Colombia. 
SDO is a mission for NASA's Living With a Star program. \facility{SDO (HMI).}

\begin{appendix}

\section{Determining the temperature and area of spatially unresolved flares}\label{sec:appendix_temp}
Assuming one observes a spatially unresolved enhancement during flares, how can we estimate the flare area and temperature? Previous studies assumed a flare area to calculate the flare temperature \citep[e.g.,][]{Kowalski2017ApJ}, or omitted the quiet-Sun luminosity from the calculation \citep[e.g.][]{Hawley2003ApJ}, which is justified for M-dwarfs, but would be less accurate for solar-type stars. The flare temperature obviously depends on the assumed area, and larger areas return cooler flare temperatures. A recalculation was already done by \citet[][]{Kowalski2017ApJ} who revised the temperature derived by \citet{Kretzschmar2011} from $\sim$9000 to $\sim$7000 K by including the quiet-Sun luminosity, but assuming an area. Similar calculations, mostly also by fixing areas to single pixels, calculated flare temperatures of 5000-7000 K \citep[e.g.][]{watanabe2013ApJ...776..123W, Kerr2014, Kleint2016ApJ...816...88K, Kowalski2017ApJ, Namekata2017ApJ...851...91N}. Here we calculate the luminosity analytically, assuming that the flare shows a black-body type spectrum, which has been observed to be valid in the visible in several cases for solar flares \citep[e.g.,][]{Kerr2014, Kleint2016ApJ...816...88K} and often their optically-thin contribution is lower \citep[e.g.][]{heinzel2017ApJ...847...48H}. We assume a pre-flare temperature, which in the solar case equals to $T_{\text{star}}=T_{\sun}=5778$\,K. This can, of course, be adapted depending on the assumption where the flare is taking place (e.g. sunspot, penumbra). During the flare, we have the area ($A_{\text{star}}-A_{\text{flare}}$) emitting at the original temperature, and a flare area ($A_{\text{flare}}$) emitting at a higher temperature ($T_{\text{flare}}$).
The pre-flare luminosity is given by  
\begin{equation}
L_{\text{pre-flare}}=\pi\int B_{\lambda}\left(T_{\text{star}}\right) A_{\text{star}} d \lambda,
\end{equation}
and during the flare
\begin{equation}
L_{\text{flare}}=\pi\int B_{\lambda}\left(T_{\text{star}}\right)\left(A_{\text{star}}-A_{\text{flare}}\right) d \lambda+\pi\int B_{\lambda}\left(T_{\text{flare}}\right) A_{\text{flare}} d \lambda.
\end{equation}
$B_{\lambda}(T)$ is the Planck function. The difference in emission during a flare is calculated as follows
\begin{equation}
\Delta L=L_{\text{flare}}-L_{\text{pre-flare }}=\pi\int B_{\lambda}\left(T_{\text{flare}}\right) A_{\text{flare}} d \lambda-\pi\int B_{\lambda}\left(T_{\text{star}}\right) A_{\text{flare}} d \lambda.
\end{equation}
To obtain the temperature of the flare, we take the ratio of the color enhancements between two different filters. To compare with previous results, let us assume the blue and green filters used by \citet{Kretzschmar2011} at 4020\,\AA{} and 5000\,\AA{}, although the following derivation is independent of the selected filter. The ration of the enhancement is
\begin{equation}\label{eqa4}
\frac{\Delta L_{\text{4020\,\AA{}}}}{\Delta L_{\text{5000\,\AA{}}}}=\frac{B_{\text{4020\,\AA{}}}\left(T_{\text{flare}}\right) \cancel{A}_{\text{flare}}-B_{\text{4020\,\AA{}}}(\text{T}_{\sun}) \cancel{A}_{\text{flare}}}{B_{\text{5000\,\AA{}}}\left(T_{\text{flare}}\right) \cancel{A}_{\text{flare}}-B_{\text{5000\,\AA{}}}(\text{T}_{\sun}) \cancel{A}_{\text{flare}}}.
\end{equation}
Note that the flare area drops out, thus allowing us to estimate the blackbody temperature of the flare without assuming any $A_{\text{flare}}$ under the condition that the observed enhancement in the two selected filters is a blackbody-type enhancement and not e.g. due to recombination continua. 

Knowing the temperature, the fractional flare area ($A_{\text{flare}}/A_{\text{star}}$), can be calculated when dividing the luminosity of the flare by the pre-flare luminosity such as
\begin{equation}\label{eqA5}
\frac{L_{\text{flare}}}{L_{\text{pre-flare }}}=\frac{B_{\lambda}\left(T_{\text{star}}\right)\left(A_{\text{star}}-A_{\text{flare}}\right)+B_{\lambda}\left(T_{\text{flare}}\right) A_{\text{flare}}}{B_{\lambda}\left(T_{\text{star}}\right)A_{\text{star}}} =1+\frac{B_{\lambda}\left(T_{\text{flare}}\right)}{B_{\lambda}\left(T_{\text{star}}\right)}\frac{A_{\text{flare}}}{A_{\text{star}}}-\frac{A_{\text{flare}}}{A_{\text{star}}}
\end{equation}

Here $\lambda$ denotes one of the selected filter wavelengths and the luminosity must be evaluated at the same wavelength.
To convert the fractional area $(A_{\text{flare}}/A_{\text{star}})$ into arcsec$^2$, we multiply by the projected area of the star $(\pi R_{\sun}^2)$. In the solar case, $R_{\sun}\approx960\arcsec$. This method uniquely determines the temperature and area of a flare for solar and stellar cases if observations in (at least) two wavelength bands exist. The re-calculated Sun-as-a-star values of the flare area and its temperature from \citet{Kretzschmar2011} are listed in table~\ref{tab:kretz2011}, and a comparison with our spatially-resolved values is shown in Figure~\ref{fig:areachanges}. This method is also applicable to stellar (super)flares. If a stellar flare is observed in two different passbands in the visible (where we assume the WL to follow blackbody radiation), for example by two different satellites, or as a spectrum, one can derive its average temperature and area independently.

\section{Occurence of WL flares}\label{sec:occurrencewlf}
Summary of the WL flares depending on the GOES classification found in the literature. Percentages should be analyzed carefully due to the low-number statistics and the different type of observations, instruments, and missions involved. The flare sample of some studies also overlapped. 

\begin{table*}[hbp!]
\centering
\begin{tabular}{llllcl}
\toprule[1.5pt]
 & \multicolumn{1}{c}{\multirow{2}{*}{X}} & \multicolumn{1}{c}{\multirow{2}{*}{M}} & \multicolumn{1}{c}{\multirow{2}{*}{C}}  & GOES class & \multicolumn{1}{c}{\multirow{2}{*}{Mission}} \\ 
& & &  & range  &  \\
\midrule[1.5pt]

\protect\citet{Neidig1983SoPh}
 & 70.8\% (17/24) &   \multicolumn{1}{c}{$-$}  &  \multicolumn{1}{c}{$-$}  & X2$>$  & SPO/MBP$^a$ \\ 

\protect\citet{Sakurai1992PASJ}
 & 100\% (3/3) &  \multicolumn{1}{c}{$-$}  &  \multicolumn{1}{c}{$-$}  & [X10,X12]   & SFT/G-band$^b$ \\ 

\protect\citet{Matthews2003}
 & 100\% (5/5) & 47.8\% (22/46)  & 12.5\% (1/8)  & [C7.8,X6.1]   & Yohkoh/SXT \\ 

\protect\citet{Wang2009RAA}
 & 100\% (3/3) & 75.0\% (3/4)  & 16.7\% (1/6)  & [C5.2,X9.0]   & Hinode/SOT \\ 
\protect\citet{Buitrago-Casas2015SoPh}
 & 60\% (6$^c$/10) & 47.8\% (22$^c$/46)  & 5.3\% (1$^c$/19)  & [C1.7,X2.2]   & SDO/HMI \\ 

\protect\citet{Kuhar2016}$^d$
 & 100\% (15/15) & \multicolumn{1}{c}{$-$}  &  \multicolumn{1}{c}{$-$}  &  [X1.1$^d$,X2.8]  & SDO/HMI \\ 

\protect\citet{Huang2016RAA}
 & 70.0\% (7/10) & 40.0\% (6/15)  &  \multicolumn{1}{c}{$-$}  & [M5.1,X3.1]  & SDO/HMI \\ 

\protect\citet{Watanabe2017ApJ}
 & 68.8\% (11/16) & 47.7\% (38/85)  &  \multicolumn{1}{c}{$-$}  & [M1.0,X3.1]  & Hinode/SOT \\ 
 
\protect\citet{Song2018AA613A69S}
 &  \multicolumn{1}{c}{$-$} &  52.2\% (12/23)$^e$  & 17.0\% (8/47)$^e$   & [C1.2,M5.6]$^e$  & SDO/HMI \\ 

\citet{Song2018ApJ...867..159S} &100\% (8/8) & 58.8\% (20/34)  & 4.3\% (2/47)  & [B6.6,X3.3]  & SDO/HMI     \\ 

Castellanos Dur\'an \& Kleint (2020) &94.4\% (17/18)  & 64.9\% (24/37)  &10.5\% (2/19)  & [B6.2,X6.9]  & SDO/HMI     \\ 

\cline{2-4}
\multicolumn{1}{r}{\textbf{All}} &  \textbf{82.1\% (92/112)}& \textbf{50.7\% (147/290)} & \textbf{10.3\% (15/146)} & &\\
\bottomrule[1.5pt]
\end{tabular} 
\caption{Comparison of the WLF occurrence depending on the GOES classification between different works.
Studies that focus on WLFs without presenting the percentage of non-WLFs are excluded. Notice that some events might overlap between the studies. However, since the methods might to detect WL emission might differ considerably, we assume that the flare sample is choosen randomly.  
$^a$Sacramento Peak Observatory/Multiband Polarimeter \citep{Neidig1983S&T....65..226N}.
$^b$ The Solar Flare Telescope \citep{Ichimoto1991}.
$^c$ The GOES classes of the WL flares are not explicity written in the paper, but they were obtained by private communacation.
$^d$\citet{Kuhar2016} did not report how many M-class flares were non-WLFs.  $^e$ The GOES class of the events were deduced using their Figure 2a and comparing it with the list of flares reported by GOES of the AR 11515 between (E45$^{\circ}$-W45$^{\circ}$). }\label{tab:wlf}
\end{table*}

\end{appendix}

\bibliography{references}
\end{document}